\documentstyle[aps,prb,eqsecnum,amsmath,amsfonts,amssymb,graphicx,multicol]
{revtex}
\newcommand{\nc}{\newcommand}
\nc{\sotimes}{\mathop{\otimes}_{s}}
\nc{\rnc}{\renewcommand}
\nc{\be}{\begin{equation}}
\nc{\ee}{\end{equation}}
\nc{\bea}{\begin{eqnarray}}
\nc{\eea}{\end{eqnarray}}
\nc{\bean}{\begin{eqnarray*}}
\nc{\eean}{\end{eqnarray*}}
\nc{\ba}{\begin{array}{@{\,}ll}}
\nc{\ea}{\end{array}}
\nc{\lt}{\left\{}
\nc{\rt}{\right.}
\nc{\nn}{\nonumber}
\nc{\mb}{\mbox}
\nc{\Y}{{\cal Y}}
\nc{\T}{{\cal T}}
\nc{\K}{{\cal K}}
\nc{\st}{{\cal S}}
\nc{\xxz}{${XXZ}$}
\nc{\dl}{\widehat{dl}}
\nc{\np}{\nonumber\\}
\nc{\mfc}{{\mathfrak{c}}}
\nc{\mfC}{{\mathfrak{C}}}
\nc{\mfa}{{\mathfrak{a}}}
\nc{\mfA}{{\mathfrak{A}}}
\nc{\mfb}{{\mathfrak{b}}}
\nc{\mfB}{{\mathfrak{B}}}

\begin{document}
%
\draft
\title{
Commuting quantum transfer matrix approach to 
intrinsic Fermion system: Correlation length
of a spinless Fermion model}
\author{Kazumitsu Sakai,
Masahiro Shiroishi and Junji Suzuki}
\address{Institute of Physics, University of Tokyo,
Komaba 3-8-1, Meguro-ku, Tokyo 153-8902, Japan}
\author{Yukiko Umeno}
\address{Department of Physics, University of Tokyo,
Hongo 7-3-1, Bunkyo-ku, Tokyo 113-0033, Japan}
\date{February 22, 1999}
\maketitle
%
%
\begin{abstract}
The quantum transfer matrix (QTM) approach to
integrable lattice Fermion systems is presented.
As a simple case we treat the spinless Fermion model with
repulsive interaction in critical regime.
We derive a set of non-linear integral equations 
which characterize the free energy and the 
correlation length of $\langle c_j^{\dagger}c_i\rangle$
for arbitrary particle density at any finite temperatures.
The correlation length is determined by solving the
integral equations numerically. 
Especially in low temperature limit
this result agrees with the prediction from conformal field theory 
(CFT) with high accuracy.
\end{abstract}
\pacs{PACS numbers: 05.30.-d, 71.10.Fd} 
\begin{multicols}{2}
\narrowtext
\section{Introduction}
%
Exact evaluations of physical quantities at finite temperatures 
pose serious difficulties even for integrable models.
One has to go much beyond mere diagonalization of a Hamiltonian;
summation over the eigenspectra must be performed.

The string hypothesis \cite{YY,G,T,TS}
brought the first breakthrough and success.
It yields a systematic way to evaluate several bulk quantities
including specific heats, susceptibilities and so on.

More recently, the quantum transfer matrix (QTM)
method has been proposed to overcome some difficulties, to which 
the standard approach is not applicable.
\cite{MSuzPB,InSuz,InSuz2,Koma,SAW,SNW,DdV,TakQT,Mizu,Klu,KZeit,JK,JKStJ,JKStp,JKSfusion,JKSHub,KSS,Klu2,FKM,FKM2,JSuz1,JSuz2,JSuz3,Sch}
One reduces the original problem
to finding the largest eigenvalue of the QTM which acts on a
fictitious system of size $N$
(referred to as the Trotter number),
which should be sent $N\to\infty$
(Trotter limit).
As this procedure is sometimes 
difficult,
we integrate its procedure with another ingredient,
the integrable structure of the underlying model.
It allows for introduction of 
the commuting QTMs 
which are labeled by complex
parameter $x$.\cite{Klu}
A set of 
auxiliary functions, including the
QTM itself,
satisfy certain functional relations.
We shall choose these functions
such that they 
have a nice analytical property
called ANZC (Analytic,
NonZero, and Constant asymptotics, see Sec.~III)
in a certain strip on the complex $x$ plane.
This admits the transformation of the functional
relations into a closed set of the integral equations.
For all cases known up to now,
the Trotter limit $N \to \infty$
can be taken analytically in the integral equations.
We thus have seen a remarkable reduction
from the problem of combinatorics (summation over the
eigenspectra)
to the study of analytic structures of 
suitably chosen auxiliary functions. 

This novel scenario has been applied to many models 
of physical interest.
\cite{Klu,KZeit,JK,JKStJ,JKStp,JKSfusion,JKSHub,KSS,Klu2,FKM,FKM2,JSuz1,JSuz2,JSuz3}
Especially, 
the correlation lengths, of which calculation
have been one of the major difficulty in the string hypothesis, 
are explicitly evaluated in the spin models.
For example of the success, we refer to recent analysis 
on the  quantum-classical crossover 
phenomena in the massless ${XXZ}$ model in ``attractive" regime.
\cite{FKM,FKM2}

We extend these studies to lattice Fermion systems.
Our formulation is fully general for the 
1D Fermion systems which are 
integrable in the sense of  the Yang-Baxter (YB) equation.
As a concrete example, we take the spinless 
Fermion model with repulsive interactions in
gapless regime.
This simple example already manifests 
some fundamental differences from the spin models,
and yields a sound basis for the future studies on more 
realistic Fermion systems such as the Hubbard model.

As in Refs.~\onlinecite{JKStJ,JKStp,JKSfusion,JKSHub},  
first one may 
perform the Jordan-Wigner (JW)
transformations to the Fermion models and 
further convert the resultant quantum spin models 
into 2D classical vertex models.
These procedures have been successful in studies of the
bulk quantities.
In evaluating correlation lengths, however, this is no longer true.
As an example, which will be discussed in the main body of this paper, 
let us take Fermion one-particle Green's function $\langle c^{\dagger}_j 
c_i \rangle$
and its correspondent
$\langle \sigma^{+}_j \sigma^{-}_i \rangle$ in the spin model.
Obviously they are related, but quite different by 
nonlocal terms due to the JW transformation.
At zero temperature ($T=0$),
using the conformal mapping, one
evaluates the scaling dimensions from the finite size 
corrections to the energy spectra. 
As the Hamiltonians are equivalent through the
JW transformations,
it is normally difficult to discriminate
between the energy spectra of
the Fermions and those of the spins.
The difference lies only
in the boundary conditions.
Nevertheless, even after JW transformation
one can explicitly calculate 
the correct scaling dimensions only by 
incorporating the proper Fermion statistics
at the very last stage (see Appendix B).
At finite temperature ($T>0$),
the QTM approach gives the correlation
function in the spectral decomposition form
as $\sum_k |A_k|^2 (\Lambda_k /\Lambda_1)^{x}$.
Here $\Lambda_k$ denotes the $k$-th largest eigenvalue of 
the QTM and $A_k$ is a certain matrix element.
%
%
Once the JW transformation is performed, 
it is difficult to trace
the difference in the statistics in this framework.
Then one hardly recognizes the difference in
the eigenvalues of the QTM 
between the spin models and the Fermion models.
A simple prescription is not yet found in contrast to the 
above mentioned case at $T=0$. 

Let us recall the quantum inverse scattering method 
based on the graded YB relation, \cite{PZ,OWA} by which
 the integrability and other algebraic
structures of the Fermion systems have been discussed successfully.
\cite{PZ,OWA,OW,MG,FM,SUW,MR}

The formulation, however, has a severe problem  in applying 
to the finite temperature case.  We must treat the quantum 
and the auxiliary spaces on the same footing when constructing 
the QTM.
On the contrary, in the graded YB relation  the quantum space is the 
Fermion Fock space, while the auxiliary space is the (graded)
vector space. 

To overcome these difficulties, 
we adopt another approach to the Fermion systems, 
which was invented quite recently. \cite{DS,USW1,USW2} 
In this method, we consider an ${R}$-operator consisting of 
the Fermion operators alone, together with its ``super-transposition".
This time both quantum and auxiliary spaces are Fermion Fock spaces. 
Therefore we can, for instance, exchange their roles with no difficulty.
Actually, by careful introduction of the super-trace and interchange of
it with the normal trace to the partition function, 
we can derive the commuting QTM for the Fermion systems.

The resultant QTM preserves genuine Fermion statistics.
In other words, the selection rule is already built-in algebraically.
This proper treatment for the statistics results in
a change of the analytic structure for the QTM. 
In the ``physical strip", the QTM has only one
additional zero which characterizes ``excited free
energy" 
at finite $T$, while in the corresponding spin model
there appear two such zeros.
Consequently, one observes a $T$-dependent
oscillating behavior of one-particle Green's function,
as well as
the difference in the correlation length between
the Fermion model and the corresponding spin model.
These are  smoothly  connected to the expected values at 
the CFT limit,
$T \to 0$ (see Appendix B).

This paper is organized as follows.
In the next section, we will present the commuting QTM formulation
of the spinless Fermion model at $T>0$.
The Fermionic $R$-operator, together with
its ``super-transposition" $\tilde{R}$, play 
fundamental roles.
The analytic structure of the QTM and the auxiliary functions are 
discussed in Sec.~III, which leads to the
nonlinear integral equations (NLIE)
characterizing the correlation length.
The limit $T \to 0$ is treated analytically at 
the ``half-filling" ($n_{\rm e}=0.5$), which recovers the prediction from CFT.
We also perform  numerical investigations on NLIE and the correlation 
length for one-particle Green's function.
To our knowledge, this is the first exact computation
of the correlation 
length  for various interaction strengths, electron filling and
for wide range of temperatures.
In Sec.~IV, we comment on 
alternative forms of NLIE derived from different
choice of the auxiliary functions. 
They are akin to the standard ``thermodynamic Bethe ansatz
(TBA) equations" from the string hypothesis,
thus may be of their own interest.
Details of calculations and supplementary knowledge on CFT are
summarized in appendices.
%
\section{Commuting Quantum Transfer Matrix for the Spinless Fermion Model }
%
In this section we formulate the commuting QTM for the spinless Fermion model. 
The formulation is based on the recent developments in the study of the 
integrability of the lattice Fermion systems.\cite{DS,USW1,USW2} 
The central role is played by an operator solution of the YB equation 
called the Fermionic ${R}$-operator.   
The ``transfer matrix" can be constructed from the ${R}$-operator, 
which generates the left-shift operator, 
the Fermionic Hamiltonian and other conserved operators. 
Here and in Sec.~II A, we briefly describe the method. 

To extend the method to the finite temperature case  
utilizing the Trotter formula, 
it is necessary to look for another transfer matrix 
which generates {\it right}-shift operator and the Hamiltonian.  
In Sec.~II B, we shall argue how to construct the desired transfer matrix 
by considering the super-transposition of the ${R}$-operator.

Based on these two kinds of the transfer matrices, 
we devise the QTM  for the Fermion model in Sec.~II C.
The QTM constitutes a one-parameter commuting family, 
which is a consequence of the global YB relation. 
The YB relation also 
enables us to diagonalize the QTM  by means of the algebraic Bethe ansatz. 
The free-energy and the correlation length are expressed in terms of 
the eigenvalues of the QTM.
%
\subsection{Fermionic ${R}$-Operator}
%
We define the spinless Fermion model by the Hamiltonian
\begin{eqnarray}
{\cal H} &:=& \sum_{j=1}^{L} {\cal H}_{j,j+1} \nonumber \\
{\cal H}_{j,j+1} &:=& \frac{t}{2} 
\Big\{ c_{j}^{\dagger} c_{j+1} + c_{j+1}^{\dagger} c_{j} 
\nonumber \\
& & \ \ \ \ \ + 2 \Delta \left(n_j - \frac{1}{2} \right) 
\left(n_{j+1} - \frac{1}{2} \right)  \Big\},
\label{eq.hamiltonian} 
\end{eqnarray}
where ${c_{j}^{\dagger}}$ and ${c_{j}}$ are the Fermionic creation
and annihilation operators at the ${j}$-th site 
satisfying the canonical anti-commutation relations
\begin{equation}
\{ c_{j}, c_{k} \} 
= \{ c_{j}^{\dagger}, c_{k}^{\dagger} \} = 0, 
\ \ \{ c_{j}^{\dagger}, c_{k} \} = \delta_{jk}. \label{eq.ACR}
\end{equation}
 We assume the periodic boundary condition (PBC) 
on the Fermion operators,
\begin{equation}
c_{L+1}^{\dagger} = c_{1}^{\dagger}, \ \ \ \ c_{L+1} = c_{1}. \label{eq.pbc}
\end{equation}
The parameters ${t,\Delta}$ are real coupling constants. 
In the present paper we consider the repulsive critical 
region ${0 \le \Delta <1, \ \ 0< t}$ and introduce the parametrization
\begin{equation}
\Delta:= \cos 2 \eta, \ \ 0 < 2 \eta \le \frac{\pi}{2}. 
\end{equation}
In the subsequent sections, we shall also use the parameter ${p_0}$ defined by
\begin{equation}
p_0:= \frac{\pi}{2 \eta}.
\end{equation}
Hereafter we set ${t=1}$ for simplicity.

The model (\ref{eq.hamiltonian}) is exactly solved by the Bethe ansatz method. 
Since the Hamiltonian (\ref{eq.hamiltonian}) preserves the number of the particles, 
we can add the ``chemical potential" term without breaking the integrability
\begin{equation}
{\cal H}_{\rm chemical} :=  \mu \sum_{j=1}^{L} \left(n_{j} - 
     \frac{1}{2}\right) \label{eq.chemical}.
\end{equation}
However we consider  only the case ${\mu =0}$ for a while.

The several physical properties including the integrability of 
the Fermion model (\ref{eq.hamiltonian}) has been discussed by 
transforming it into the ${XXZ}$ model  
\begin{eqnarray}
H = \frac{1}{4} \sum_{j=1}^{L} 
\Big\{ \sigma_{j}^{x} \sigma_{j+1}^{x} + \sigma_{j}^{y} \sigma_{j+1}^{y} 
+ \Delta \sigma_{j}^{z} \sigma_{j+1}^{z}  \Big\},
\label{eq.spinhamiltonian} 
\end{eqnarray} 
through the JW transformation. 
However it was recently discovered that we can treat 
the Fermion model (\ref{eq.hamiltonian}) only with the Fermion operators. 
We shall summarize the method in what follows.  

First let us consider a two-dimensional Fermion Fock space ${V_j}$,
 a basis of which is given by
\begin{eqnarray}
& & | 0 \rangle_{j}, \ \ \ \ | 1 \rangle_{j} := c_j^{\dagger}| 0 \rangle_{j}, \nonumber \\ 
& & c_j | 0 \rangle_{j} = 0.
\end{eqnarray}

Define the Fermionic ${R}$-operator acting on the tensor product of 
the Fermion Fock spaces ${V_j {\displaystyle \sotimes} V_k}$ by
\begin{eqnarray}
{\cal R}_{jk}(v) &:=&  
a(v) \left\{ - n_j n_k 
+ (1-n_j)(1-n_k) \right\} \nonumber \\ 
& & + b(v) \left\{ n_j (1-n_k) + (1-n_j) n_k \right\} \nonumber \\
& & + c(v) ( c_j^{\dagger} c_k -  c_j c_k^{\dagger}), 
\label{eq.fR}
\end{eqnarray}
where 
\begin{equation}
a(v):= \frac{\sin \eta (v + 2)}{\sin 2 \eta}, \ \ 
b(v):= \frac{\sin \eta v}{\sin 2 \eta}, \ \ c(v):= 1.
\end{equation}
A basis of ${V_j {\displaystyle \sotimes} V_k}$ is given by
\begin{eqnarray}
 & & | 0 \rangle_{j} \sotimes | 0 \rangle_{k} := | 0 \rangle, 
 \ \  | 1 \rangle_{j} \sotimes | 0 \rangle_{k} := c_j^{\dagger}| 0 \rangle, \nonumber \\
& & | 0 \rangle_{j} \sotimes | 1 \rangle_{k} := c_k^{\dagger}| 0 \rangle, 
\ \ | 1 \rangle_{j} \sotimes | 1 \rangle_{k} := c_j^{\dagger} c_k^{\dagger}| 0 \rangle,
\end{eqnarray}
and we can calculate the matrix elements of (\ref{eq.fR}) if necessary. 
However we keep the operator form (\ref{eq.fR}) as much 
as possible and avoid the use of the matrix elements, 
because the former is more transparent.
The ${R}$-operator (\ref{eq.fR}) satisfies the following YB equation \cite{USW1,USW2}
\begin{equation}
{\cal R}_{12}(u-v) {\cal R}_{13}(u) {\cal R}_{23}(v) 
= {\cal R}_{23}(v) {\cal R}_{13}(u) {\cal R}_{12}(u-v).
\label{eq.YBE1}
\end{equation}
The equation (\ref{eq.YBE1}) is an operator identity and 
one should carefully use the anti-commutation relations (\ref{eq.ACR}) 
to confirm its validity.  

It is one of the fundamental properties of the 
${R}$-operator ${{\cal R}_{ij}(v)}$ that 
${{\cal R}_{ij}(0) = {\cal P}_{ij}}$ is the permutation operator 
for the Fermion operators,
\begin{eqnarray}
& &  {\cal P}_{jk} := (1-n_j)(1 - n_k) - n_j n_k + c_j^{\dagger} 
c_k - c_j c_k^{\dagger}, \nonumber \\
& &  {\cal P}_{jk} \ x_{j} = x_{k}  {\cal P}_{jk},  \ \ \ \ ( x_{j} =  c_j \ 
{\rm or} \ c_j^{\dagger}). \label{eq.PPc}
\end{eqnarray}

We can define an analog of the transfer matrix by
\begin{eqnarray}
     T(v)  &:=& {\rm Str}_a \left\{ {\cal R}_{aL}(v) \cdots {\cal R}_{a1}(v)
 \right\}.
 \label{eq.transfer1} 
\end{eqnarray}
Here the super-trace  of an arbitrary operator ${X}$ is defined by
\begin{equation}
{\rm Str}_a X:= {}_{a} \langle 0 | X | 0 \rangle_{a}
         - {}_{a} \langle 1 | X | 1 \rangle_{a},
\label{eq.str}
\end{equation}
where the  dual Fermion Fock space is spanned by 
${ {}_{a} \langle 0 | }$ and ${ {}_{a} \langle 1 |  }$ with 
\begin{equation}
\hspace{5mm} {}_a \langle 0 |c_a^{\dagger} = 0, 
\ \ \hspace{5mm} {}_{a} \langle 1 |:= {}_a \langle 0 |c_a. 
\end{equation}
We also assume 
\begin{equation}
{}_a \langle  0 | 0 \rangle_a
= {}_a \langle  1 | 1 \rangle_a=1.
\end{equation}
The  super-trace (\ref{eq.str}) corresponds to 
the PBC for the Fermion operators (\ref{eq.pbc}) satisfies the property 
\begin{eqnarray}
& & {\rm Str}_a \left\{ {\cal R}_{aL}(v) \cdots {\cal R}_{a1}(v)
 \right\} \nonumber \\
& & =  {\rm Str}_a \left\{ {\cal R}_{a1}(v) {\cal R}_{aL}(v) \cdots {\cal R}_{a2}(v) \right\}.
\end{eqnarray}
Hereafter we call (\ref{eq.transfer1}) the transfer matrix for simplicity.

As in the case with the integrable spin models, 
the YB equation (\ref{eq.YBE1}) ensures 
the commutativity of the transfer matrices (\ref{eq.transfer1})
\begin{equation}
\left[ T(v), T(v') \right] = 0.
\end{equation}

The expansion of the transfer matrix (\ref{eq.str}) with respect to 
the spectral parameter ${v}$ is given by
\begin{equation}
T(v) = T(0) \left\{ 1 + \frac{2 \eta}{\sin 2 \eta} \left( {\cal H} +
 \frac{L}{4} \Delta \right) v + {\cal O}(v^2) \right\},
\label{eq.expansion1}
\end{equation}
which follows from the relationship
\begin{eqnarray}
& & \frac{{\rm d} {\cal R}_{aj}(v)}{{\rm d} v} \Big|_{v=0} {\cal P}_{a,j-1} \nonumber \\
& & = \frac{2 \eta}{\sin 2 \eta} {\cal P}_{aj} {\cal P}_{a,j-1} \left( {\cal H}_{j-1,j} + \frac{1}{4} \Delta \right).
\end{eqnarray} 
Note that the operator 
${T(0)={\rm Str}_a \{ {\cal P}_{aL} \cdots {\cal P}_{a1} \} }$ 
is the left-shift operator
\begin{equation}
T(0) x_{j} = x_{j+1} T(0), 
\ \ \ \ \ \ ( x_{j} =  c_j \ {\rm or} \ c_j^{\dagger}). \label{eq.left}
\end{equation}
One can easily prove the relation (\ref{eq.left}) 
utilizing the property of the permutation operator,
\begin{equation}
{\cal P}_{a,j+1}  {\cal P}_{aj} \ x_{j} = x_{j+1}  {\cal P}_{a,j+1}  {\cal P}_{aj}, 
 \ \ \ \ ( x_{j} =  c_j \ {\rm or} \ c_j^{\dagger}). 
\end{equation}
%
\subsection{ Super-Transposed Fermionic ${R}$-Operator}
%
In this section, we shall consider another transfer matrix 
which generates the right-shift operator.  
For this purpose we first define the 
super-transposition ${{\rm st}_j}$ for 
an arbitrary operator ${X_{j}(v)}$ in the form
\begin{equation}
X_j(v) = A(v)(1-n_j) + D(v) n_j + B(v) c_j + C(v) c_j^{\dagger},
\end{equation}
by
\begin{equation}
X_j^{{\rm st}_j}(v):= A(v)(1-n_j) + D(v) n_j + B(v) c_j^{\dagger} - C(v) c_j. 
\end{equation}
Here ${A(v)}$ and ${D(v)}$ (${B(v)}$ and ${C(v)}$) are 
assumed to be Grassmann even (odd) operators. 

Now applying the super-transposition ${{\rm st}_1}$ to both sides of the YB equation 
(\ref{eq.YBE1}), we obtain
\begin{equation}
{\cal R}_{13}^{{\rm st}_1}(u) {\cal R}_{12}^{{\rm st}_1}(u-v) {\cal R}_{23}(v)
=  {\cal R}_{23}(v) {\cal R}_{12}^{{\rm st}_1}(u-v)
{\cal R}_{13}^{{\rm st}_1}(u),
\end{equation}
where we have used a property of the super-transposition
\begin{equation}
\left( {\cal R}_{jk}(u) {\cal R}_{jl}(v) \right)^{{\rm st}_j} =
{\cal R}_{jl}^{{\rm st}_{j}}(v) {\cal R}_{jk}^{{\rm st}_{j}}(u), \ \ \ \ \ \ 
(k \ne l).
\end{equation}
Then changing suffixes and  spectral parameters as
\begin{eqnarray}
& &  1 \rightarrow 3, \ \ 2 \rightarrow 1, \ \ 3 \rightarrow 2, \nonumber \\
& &  u \rightarrow -v, \ \ v \rightarrow u-v, 
\end{eqnarray}
we get the following new type of the YB equation
\begin{equation}
{\cal R}_{12}(u-v) \widetilde{\cal R}_{13}(u) 
\widetilde{\cal R}_{23}(v) 
= \widetilde{\cal R}_{23}(v) \widetilde{\cal R}_{13}(u) {\cal R}_{12}(u-v),
\label{eq.YBE2}
\end{equation}
where
\begin{eqnarray}
\widetilde{\cal R}_{jk}(v) & := &  {\cal R}_{kj}^{{\rm st}_k}(-v) \nonumber \\
& = & a(-v) \left\{ - n_j n_k 
+ (1-n_j)(1-n_k) \right\} \nonumber \\
& & + b(-v) \left\{ n_j (1-n_k) 
+ (1-n_j) n_k \right\} \nonumber \\
& & - c(-v) ( c_j^{\dagger} c_k^{\dagger} + c_j c_k ). \label{eq.tildefR}
\end{eqnarray}

Although the new ${R}$-operator ${\widetilde{\cal R}_{jk}(v)}$ is not 
symmetric (${\widetilde{\cal R}_{jk}(v) \ne \widetilde{\cal R}_{kj}(v)}$), 
it is still possible to prove the relation 
\begin{equation}
\widetilde{\cal R}_{12}(u-v) \widetilde{\cal R}_{13}(u) 
{\cal R}_{23}(v) 
= {\cal R}_{23}(v) \widetilde{\cal R}_{13}(u) \widetilde{\cal R}_{12}(u-v).
\label{eq.YBE3}
\end{equation}
Using ${\widetilde{\cal R}_{aj}(v)}$, we define another transfer matrix by
\begin{eqnarray}
     \widetilde{T}(v)&:=& 
     {\rm Str}_a \left\{ \widetilde{\cal R}_{aL}(v) \cdots \widetilde{\cal R}_{a1}(v)
 \right\}. 
 \label{eq.transfer2} 
\end{eqnarray}
Then the commutative properties of 
the transfer matrices follow from the YB equations 
(\ref{eq.YBE2}) and (\ref{eq.YBE3}),
\begin{equation}
\left[ T(v), \widetilde{T}(v') \right] 
= \left[ \widetilde{T}(v), \widetilde{T}(v') \right] = 0. \label{eq.comm1}
\end{equation}

The following remarkable relations hold
\begin{equation}
\widetilde{\cal P}_{aj} \widetilde{\cal P}_{a,j-1} x_{j}  =  
x_{j-1} \widetilde{\cal P}_{aj} \widetilde{\cal P}_{a,j-1}, 
\ \ (x_j = c_j \ {\rm  or} \ c_j^{\dagger}), 
\label{eq.barPPC}
\end{equation}
where
\begin{eqnarray}
\widetilde{\cal P}_{jk} & := & \widetilde{\cal R}_{jk}(0) \nonumber \\
& =  & (1-n_j)(1 -n_k) - n_j n_k - (c_j^{\dagger} c_k^{\dagger} + c_j c_k). 
\end{eqnarray}
Using the relations (\ref{eq.barPPC}), 
one can confirm that the operator ${\widetilde{T}(0)}$ 
provides the right-shift operator, i.e., 
\begin{equation}
 \widetilde{T}(0) x_{j} = x_{j-1} \widetilde{T}(0), 
 \ \ \ \ (x_j = c_j \ {\rm  or} \ c_j^{\dagger}). 
\end{equation}
In other words, 
${\widetilde{T}(0)}$ is the inverse of  ${T(0)}$
\begin{equation}
T(0) \widetilde{T}(0) = 1. \label{eq.inverse}
\end{equation}
Furthermore, from the relationship
\begin{eqnarray}
& & \widetilde{\cal P}_{a,j+1} \frac{{\rm d} 
     \widetilde{\cal R}_{aj}(v)}{{\rm d} v} \Big|_{v=0} \nonumber \\
& & = - \frac{2 \eta}{\sin 2 \eta} \widetilde{\cal P}_{a,j+1} 
       \widetilde{\cal P}_{aj} \left( {\cal H}_{j+1,j} + 
       \frac{1}{4} \Delta \right),
\end{eqnarray}
the expansion of the transfer matrix ${\widetilde{T}(v)}$ 
with respect to the spectral parameter ${v}$ is given  by
\begin{equation}
\widetilde{T}(v) =
 \widetilde{T}(0) \left\{ 1 - \frac{2 \eta }{\sin 2 \eta} 
 \left( {\cal H} + \frac{L}{4} \Delta \right) v + 
 {\cal O}(v^2) \right\}. \label{eq.expansion2}
\end{equation}
%
\subsection{Commuting Quantum Transfer Matrix}
%
The expansions (\ref{eq.expansion1}) and (\ref{eq.expansion2}) 
with the relation (\ref{eq.inverse}) are combined into a formula
\begin{equation}
T(u) \widetilde{T}(-u) = 1 + \frac{4 \eta }{\sin 2 \eta} 
   \left( {\cal H} + \frac{L}{4} \Delta \right) u + {\cal O}(u^2).
\end{equation}
This facilitates the investigation
the finite temperature properties of the spinless Fermion model 
(\ref{eq.hamiltonian}) via the Trotter formula,
\begin{eqnarray}
 \exp \left( - \beta \left( {\cal H} +\frac{L}{4} \Delta \right) \right) 
 &=& \lim_{N \rightarrow \infty} \left( T(u_N) \widetilde{T}(-u_N) \right)^{N/2}, 
 \nonumber \\
& & u_N = -\frac{\beta \sin 2 \eta}{2 \eta N}. 
\end{eqnarray}
Here an (even) integer ${N}$ called the Trotter number, 
represents the number of sites in the fictitious 
Trotter direction and ${\beta}$ 
is the inverse temperature ${\beta = 1/T}$. 

The free energy per site, for instance, is given by
\begin {equation}
f = - \lim_{L \rightarrow \infty} \lim_{N \rightarrow \infty} 
      \frac{1}{L \beta} \ln  {\rm Tr} 
      \left(T(u_N) \widetilde{T}(-u_N) \right)^{N/2} - \frac{1}{4} \Delta.
\label{eq.fenergy}
\end{equation}
However, as is the case with the corresponding spin model, 
the eigenvalues of ${T(u_N) \widetilde{T}(-u_N)}$ are infinitely 
degenerate in the limit ${N \rightarrow \infty}$.  

Therefore it is a formidable task to take the trace in this limit. 
To avoid this difficulty, we transform the term 
${{\rm Tr} \left(T(u_N) \widetilde{T}(-u_N) \right)^{N/2}}$ in 
(\ref{eq.fenergy}) as follows:
\begin{eqnarray}
& & {\rm Tr} \left(T(u_N) \widetilde{T}(-u_N) \right)^{N/2} = \nonumber \\
& & = {\rm Tr} \prod_{m=1}^{N/2} {\rm Str}_{a_{2m},a_{2m-1}} 
  \Big[ {\cal R}_{a_{2m},L}(u_N) \cdots {\cal R}_{a_{2m},1}(u_N) \nonumber \\
& &  \hspace{1cm}\times\widetilde{\cal R}_{a_{2m-1},L}(-u_N) 
      \cdots \widetilde{\cal R}_{a_{2m-1},1}(-u_N) \Big], \nonumber \\
& & =  {\rm Str} \ \prod_{j=1}^{L} {\rm Tr}_{j} \prod_{m=1}^{N/2} 
       {\cal R}_{a_{2m},j}(u_N) \widetilde{\cal R}_{a_{2m-1},j}(-u_N).
\end{eqnarray}

We now introduce a fundamental object in the present approach 
called the quantum transfer matrix (QTM) 
\begin{equation}
T_{\rm QTM}(u_N,v):= {\rm Tr}_j {\cal T}_{j}(u_N,v), \label{eq.QTM} 
\end{equation}
where the monodromy operator ${{\cal T}_{j}(u_N,v)}$ is defined by
\begin{equation}
{\cal T}_{j} (u_N,v):= \prod_{m=1}^{N/2} {\cal R}_{a_{2m},j}(v+u_N) 
\widetilde{\cal R}_{a_{2m-1},j}(v-u_N). \label{eq.QTMmonodromy}
\end{equation}
Using the YB equations (\ref{eq.YBE1}) and (\ref{eq.YBE2}), 
we can show that the monodromy operator satisfies the global YB relation
\begin{eqnarray}
& & {\cal R}_{21}(v-v') {\cal T}_{1}(u_N,v) {\cal T}_{2}(u_N,v') \nonumber \\
& & = {\cal T}_{2}(u_N,v') {\cal T}_{1}(u_N,v) {\cal R}_{21}(v-v').
\label{eq.QTMYBE}
\end{eqnarray} 
Accordingly the QTM constitutes a commuting family
\begin{equation}
\left[ T_{\rm QTM}(u_N,v), T_{\rm QTM}(u_N,v') \right] = 0.
\end{equation}
We remark that the trace in the definition of the QTM (\ref{eq.QTM}) 
implies the anti-periodic boundary condition for 
the Fermion operators in the Trotter direction, \cite{USW2} 
i.e., 
\begin{equation}
c_{a_{N+1}} = - c_{a_1}, \ \ c_{a_{N+1}}^{\dagger} = - c_{a_1}^{\dagger}.
\end{equation}

The free energy per site (\ref{eq.fenergy}) is then 
represented in terms of the QTM as
\begin{equation}
f = - \lim_{L \rightarrow \infty} \lim_{N \rightarrow \infty}  
\frac{1}{L \beta} \ln  {\rm Str} \left(T_{\rm QTM}(u_N,0)^{L} \right) - 
\frac{1}{4} \Delta.
\label{eq.fenergy2}
\end{equation}

Since the two limits in (\ref{eq.fenergy2}) 
are exchangeable, \cite{MSuzPB,InSuz} 
we take the limit ${L\rightarrow \infty}$ first. 
Because there is a finite gap between the first and 
the second largest eigenvalue of the QTM for finite temperature, 
we can write 
\begin{equation}
f = - \frac{1}{\beta} \lim_{N \rightarrow \infty} \ln \Lambda_1 - \frac{1}{4} \Delta.
\end{equation} 
where  ${\Lambda_1}$ is the first largest 
eigenvalue of the QTM ${T_{\rm QTM} (u_N,0)}$.
From now on $\Lambda_k$ denotes the
$k$-th largest eigenvalue of the QTM.
The correlation length ${\xi}$ of the 
correlation function ${\langle c_j^{\dagger} c_k \rangle}$ 
can also be represented in terms of the  first 
and the second largest eigenvalues $\Lambda_2$ as
\begin{equation}
\xi^{-1} = - \lim_{N \rightarrow \infty} 
\ln \Big| \frac{\Lambda_2}{\Lambda_1} \Big|. \label{eq.xi}
\end{equation}
In this way the calculation of certain thermal quantities 
reduces to the evaluation of the eigenvalues of the QTM 
in the Trotter limit (${N \rightarrow \infty}$).

For ${N}$ finite, 
it is possible to diagonalize the QTM (\ref{eq.QTM}) 
by means of the algebraic Bethe ansatz \cite{KBI}
(see Appendix A). 
The eigenvalue is then given by
\begin{eqnarray}
& & \Lambda(x) = \lambda_1(x) + \lambda_2(x), \nonumber \\
& & \lambda_1(x):=\phi_{+}(x) \phi_{-}(x - 2 i)
    \frac{Q(x + 2 i)}{Q(x)}e^{ \beta \mu /2} , \nonumber \\ 
& & \lambda_2(x):=(-1)^{N/2 + N_{\rm e}} 
     \phi_{-}(x) \phi_{+}(x  + 2 i) \frac{Q(x - 2 i)}{Q(x)}e^{ -\beta \mu /2}, 
                          \nonumber \\ \label{eq.eigen}
\end{eqnarray}
where
\begin{eqnarray}
\phi_{\pm}(x) &:=& \left( 
     \frac{\sinh \eta (x \pm i u_{N})}{\sin 2 \eta} \right)^{N/2}, \np
     Q(x) &:=& \prod_{j=1}^{N_{\rm e}} \sinh \eta (x- x_j).
\label{qdef}
\end{eqnarray}
Here we have changed the spectral parameter from 
${v}$ to ${x}$ defined by  ${v = i x}$ for later convenience.
Note that we have also included the contribution 
from the chemical potential term (\ref{eq.chemical}) 
in the expression (\ref{eq.eigen}). 

The associated Bethe ansatz equation (BAE) is given by
\begin{eqnarray}
& & \left( \frac{\phi_{+}(x) \phi_{-}(x - 2 i)}
{\phi_{-}(x) \phi_{+}(x + 2 i )} \right)^{N/2} \nonumber \\
& & = -  (-1)^{N/2 + N_{\rm e}}  e^{-\beta \mu } \prod_{k =1}^{N_{\rm e}}
\frac{Q(x_j - 2 i)}{Q(x_j + 2 i)}. \label{eq.bae}
\end{eqnarray} 
Compared with the ${XXZ}$ model, 
we observe an extra factor ${(-1)^{N/2 + N_{\rm e}}}$ 
in (\ref{eq.eigen}) and (\ref{eq.bae}) which reflects 
the Fermionic nature of the present system. 
In particular, 
if ${N/2 + N_{\rm e} \equiv 1 \ ({\rm mod} \ 2)}$, 
the Eqs. (\ref{eq.eigen}) and (\ref{eq.bae}) are 
clearly different from the corresponding ones for the ${XXZ}$ model.
Actually the second largest eigenvalue lies in the sector ${N_{\rm e} = N/2-1}$, 
while the first largest one is in the sector ${N_{\rm e} = N/2}$. 
Therefore the correlation length ${\xi}$ (\ref{eq.xi}) 
exhibits the manifest difference between the Fermion system 
(\ref{eq.hamiltonian}) and the spin system (\ref{eq.spinhamiltonian}). 
%
%
\section{NLIE and the Exact enumeration of correlation length}
%
\subsection{Analyticities of Auxiliary Functions and NLIE}
%
In order to proceed further, one needs to
clarify the analytic property of the QTM.
For this purpose, we perform numerical investigations
by fixing the Trotter number $N$ finite.

First we give the description for the largest eigenvalue 
sector, which is naturally identical to the corresponding
$XXZ$ model.
There are $N_{\rm e} = N/2$ BAE roots.
Only at the ``half-filling", they distribute exactly on the real axis
symmetrically with respect to $x=0$, while for the general
particle density they bend in the complex $x$ plane.
The QTM has $N$ zeros in $\Im x\in[-2p_0,2p_0]$:
$N/2$ zeros locate on the smooth curve $ \Im x \sim 2$, 
and the other $N/2$ zeros are on the curve $ \Im x \sim -2$.
Thus there is a  strip $ \Im x\in [ -1, 1 ]$ where
the QTM is analytic and nonzero. We call this ``physical strip".

Next consider the excited state relevant to the
second largest eigenvalue.
In contrast to the $XXZ$ model, we find that  
two complex eigenvalues are degenerate in magnitude.
Both of them are characterized by
$N_{\rm e} = N/2-1$ BAE roots located on a smooth curve
near the real axis. 
The distribution of the BAE roots for the one and
that for the other are symmetric with respect to
the imaginary axis.
As to the zeros of the QTM, $N-2$ zeros are on the
smooth curves $ \Im x \sim \pm 2$.

The locations of the two
``missing zeros" are vital in the evaluation of the 
excited states.
For the $XXZ$ model, both of them enter into the physical strip.
Especially, with vanishing external field $h$, 
they are on the real axis and are symmetric with respect
to the imaginary axis.
With the increase of $h$, they are away from the real
axis, but still stay in
the physical strip preserving the symmetry.

\begin{figure}
\begin{center}
\includegraphics[width=0.48\textwidth]{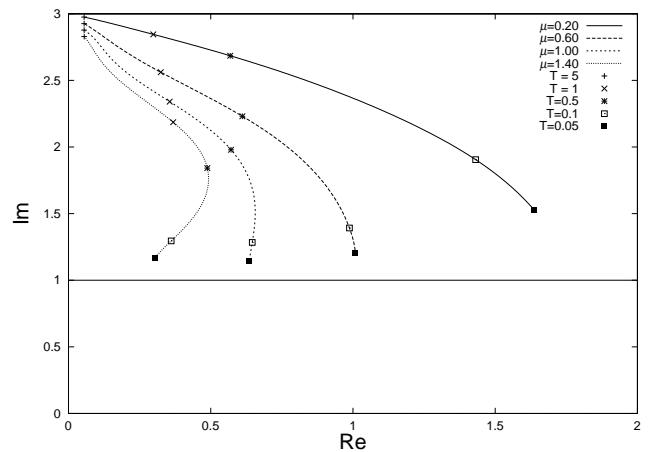}
\end{center}
\caption{The trajectories of the 
additional zero $\theta'$
are depicted in the case $p_0=3$, $N=100$.
With the decrease of $T$, $\theta'$ moves
downward, whereas it never
comes into the physical strip.}
\label{zero}
\end{figure}

We find a different situation for the Fermion model.
At the half-filling, corresponding to $h=0$ in the $XXZ$ model, 
one of them is located at $\theta_0$ on the real axis, 
while the other is at $\theta' _0+ i p_0$, 
$\theta _0\sim \theta'_0$.
%
%
Namely only one zero appears in the physical strip.
Away from the half-filling, the zero in the physical strip
(we call it $\theta$) moves upward while 
the other ($\theta'$) moves downward.
Nevertheless, we find that $\theta$ remains in the physical strip
while $\theta'$ never comes in.
{}From now on we consider the case $\Re\theta>0$ ($\Re\theta'>0$).
Then the trajectories of $\theta'$, for example,
are depicted in FIG.~\ref{zero}. 

We assume all these features are valid in the Trotter limit
$N \rightarrow \infty$.
Then a set of nonlinear integral equations (NLIE) 
can be derived
as in the case of the $XXZ$ model.\cite{KZeit}
We define auxiliary functions 
\begin{eqnarray}
{\mathfrak{a}}(x) &:=& 
\frac{\lambda_1(x+i-i\gamma_1)}{\lambda_2(x+i-i\gamma_1)},\qquad 
{\mathfrak{A}}(x):= 1+{\mathfrak{a}}(x), \np
\bar{{\mathfrak{a}}}(x) &:=& 
\frac{\lambda_2(x-i+i\gamma_2)}{\lambda_{1}(x-i+i\gamma_2)}, \qquad 
\overline{{\mathfrak{A}}}(x):= 1+\overline{{\mathfrak{a}}}(x). 
\label{bdef}
\end{eqnarray}
where $\gamma_1, \gamma_2$ are small positive quantities introduced
for the convenience in numerical calculations.
Note that these functions have asymptotic values
\begin{subequations}
\bea
\mfa(x)&=&\begin{cases}
     \exp((-\pi+4\eta)i+\beta\mu)  & \text{for $x\to-\infty$,} \\
     \exp((\pi-4\eta)i+\beta\mu) & \text{for $x\to\infty$,} 
     \end{cases}  \label{asymptotics1} \\
\overline{\mfa}(x)&=&\begin{cases}
     \exp((\pi-4\eta)i-\beta\mu)  & \text{for $x\to-\infty$,} \\
     \exp((-\pi+4\eta)i-\beta\mu) & \text{for $x\to\infty$.} 
      \end{cases}
      \label{asymptotics2}
\eea
\end{subequations}
Immediately seen from the above analyticity argument,
${\mathfrak{a}}(x),{\mathfrak{A}}(x)$
($\overline{{\mathfrak{a}}}(x),\overline{{\mathfrak{A}}}(x)$) are Analytic,
NonZero and have Constant asymptotic values (ANZC) in a certain
strip  in the lower (upper) half plane including real axis.
The above definitions, together with the 
knowledge of zeros for $\Lambda(x)$,
fix the  NLIE among these auxiliary functions.
We defer the detail derivation 
to Appendix C.
The resultant expressions allow for taking the Trotter limit 
analytically.
Thereby one arrives at the final expressions totally independent
of fictitious parameter $N$,
\begin{eqnarray}
   \ln {\mathfrak{a}}(x) &=& 
     -\frac{\pi \beta  \sin2\eta}{4\eta \cosh\frac{\pi}{2}(x-i\gamma_1)}+
      F*  \ln {\mathfrak{A}}(x) \np
    &&- F* \ln \overline{{\mathfrak{A}}}(x+2i-i(\gamma_1+\gamma_2)) \np
    &&+2\pi i {\cal F}(x-\theta+i(1-\gamma_1))  
    +\frac{\beta \mu p_0}{2(p_0-1)},  \np
\ln \overline{{\mathfrak{a}}}(x) &=& 
     -\frac{\pi \beta  \sin2\eta}{4\eta \cosh\frac{\pi}{2}(x+i\gamma_2)}+
      F* \ln \overline{{\mathfrak{A}}}(x) \np
    &&-F* \ln {\mathfrak{A}}(x-2i+i(\gamma_1+\gamma_2)) \np
    &&-2\pi i {\cal F}(x-\theta-i(1-\gamma_2))  
    -\frac{\beta \mu p_0}{2(p_0-1)}.  
\label{NLIE}
\end{eqnarray}
where 
\begin{eqnarray}
  &&A*B(x):=\int_{-\infty}^{\infty} A(x-y) B(y) dy,   \np
&&F(x):= \frac{1}{2\pi} \int_{-\infty}^{\infty} 
             \frac{\sinh(p_0 -2) k}{2  \cosh k \sinh(p_0-1)k } 
			 e^{-ikx} dk, \np
&&{\cal  F} (x):=\frac{i}{2\pi} \int_{-\infty}^{\infty} 
             \frac{\sinh(p_0 -2) k}{2k \cosh k \sinh(p_0-1)k } 
			 e^{-ikx}dk. 
\label{DefF}
\end{eqnarray}
The location of zero $\theta$ satisfies a subsidiary condition,
\begin{equation}
{\mathfrak{a}}(\theta-i+i\gamma_1) =-1.
\label{sub}
\end{equation}
Taking the Trotter limit $N\to\infty$ after setting $x=0$ in
(\ref{second})
we derive the ``first excited free energy " 
per site $f_2$ is
\begin{eqnarray}
f_2&&=-\frac{1}{\beta} \ln \Lambda_2 (0)-\frac{1}{4}{\Delta} \np 
 &&=\epsilon_0 -\frac{1}{\beta} K* \ln{\mathfrak{A}}(i\gamma_1) -
 \frac{1}{\beta} K* \ln \overline{{\mathfrak{A}}}(-i\gamma_2),\np
 &&-\frac{1}{\beta} \ln \tanh\frac{\pi \theta}{4}
        -i\frac{\pi}{2\beta}
\label{sec}
\end{eqnarray}
where $\epsilon_0$ is the ground state
energy defined in (\ref{ground}) and
\be
  K(x) := \frac{1}{4\cosh \frac{\pi x}{2} }.
\end{equation}

Together with the NLIE for the largest eigenvalue, 
summarized in Appendix C, 
these relations characterize the correlation length $\xi$ of
one-particle Green's function $\langle c^{\dagger}_j,c_i \rangle$
at $T>0$ completely (see Eq.~(\ref{eq.xi})).

We remark that in derivations of above relations
one does not need precise information like 
roots distributions of the BAE.
Only ANZC properties of the QTM and the auxiliary functions
are sufficient.
Thus the structure is rather robust, and permits to introduce
small free parameters
$\gamma_1$ and $\gamma_2$.

In the next two subsections, we present analytical and numerical
studies on these equations and the correlation length 
of one-particle Green's function, 
which are main results in this paper.

\subsection{Low temperature property of NLIE ($\mu=0$)}

We study the low temperature behavior for the
half-filling case $\mu=0$ utilizing
the Dilogarithm trick \cite{KP}, which enables us to 
obtain the first low temperature correction without solving NLIE.
As in the case of the largest eigenvalue sector,
$|\mfa (x)|$ and $|\overline{\mfa}(x)|$ exhibit a crossover behavior,
\be
\begin{cases}
  |\mfa(x)|,|\overline{\mfa}(x)|
     \ll 1 &\text{for $|x|<{\cal K}$,} \\
  |\mfa(x)|,|\overline{\mfa}(x)|
     \sim 1 &\text{for $|x|>{\cal K}$,}
   \end{cases}
\label{crossover}
\end{equation}
where 
\be
{\cal K}:=
\frac{2}{\pi}\ln \frac{\pi\beta\sin(2\eta)}{2\eta}.
\end{equation}
Thus one carefully takes into account of
contributions near ``Fermi-surfaces" $\pm{\cal K}$.
For this purpose, we introduce following 
shifted variables and scaling functions,
\begin{eqnarray}
la_{\pm}(x) &:=& \ln {\mathfrak{a}}\left(\pm \frac{2}{\pi}x
                  \pm {\cal K}\right), \np
l\overline{a}_{\pm}(x) &:=& \ln \overline{{\mathfrak{a}}}\left(
                 \pm\frac{2}{\pi}x\pm {\cal K}\right), \np
\overline{\theta} &:=& \frac{\pi}{2}(\theta - {\cal K}).
\label{scaling}
\end{eqnarray}
and similarly for capital functions 
$\mfA$, $\overline{\mfA}$, $A_{\pm}$ and $\overline{A}_{\pm}$.
In $T \rightarrow 0$, they satisfy truncated equations,
\begin{subequations}
\begin{eqnarray}
la_{+}(x) &=&  -e^{-x +\frac{\pi}{2}i\gamma_1}+
               F_1*lA_{+}(x) -F_2*l\overline{A}_{+}(x)  \np
& & +2\pi i {\cal  F}\left(\frac{2}{\pi}(x - \overline{\theta}) + 
                        i (1 - \gamma_1)\right), 
\label{scaleNLIE1} \\
l\overline{a}_{+}(x) &=&-e^{-x-\frac{\pi}{2}i\gamma_2}+
  F_1*l\overline{A}_{+}(x) -\overline{F}_2*lA_{+}(x)  \np
& & - 2\pi i {\cal  F}\left(
     \frac{2}{\pi}(x-\overline{\theta})-i(1-\gamma_2)\right),  
\label{scaleNLIE2}  \\
la_{-}(x) &=& -e^{-x-\frac{\pi}{2}i\gamma_1} +
      F_1*lA_{-}(x) - \overline{F}_2*l\overline{A}_{-}(x) \np
& & + 2\pi i {\cal  F}(- \infty), 
\label{scaleNLIE3}   \\
l\overline{a}_{-}(x) &=& -e^{-x + \frac{\pi}{2}i\gamma_2} +
      F_1*l\overline{A}_{-}(x) -F_2*lA_{-}(x) \np
& & - 2\pi i {\cal  F}(- \infty), 
\label{scaleNLIE4} 
\end{eqnarray}
\end{subequations}
where 
\bea
F_1(x)&&:=\frac{2}{\pi} F\left(\frac{2 x}{\pi}\right),\np 
F_2(x)&&:=\frac{2}{\pi} F\left(\frac{2}{\pi}x+2i-
            i(\gamma_1+\gamma_2)\right).
\eea
and $\overline{F}_1, \overline{F}_2$ are their complex conjugate.
In this limit, the finite $T$ correction part, $\ln\Lambda_{\rm fn}(x)$ 
(see (\ref{finiteT})) reads 
\begin{eqnarray}
\ln&&\Lambda_{\rm fn}(x) \sim 
  \frac{\pi}{2}i+\frac{2\eta}{\pi^2 \beta \sin 2 \eta}
    \Bigl (  
       -2\pi e^{\frac{\pi}{2}x - \overline{\theta}} \np
& & + e^{\frac{\pi}{2}x}\int_{-\infty}^{\infty} e^{-y} 
   \{ e^{\frac{\pi}{2}i\gamma_1} lA_{+}(y) + 
    e^{-\frac{\pi}{2}i\gamma_2} l{\overline A}_{+}(y)\} dy    \np      
 & & +e^{-\frac{\pi}{2}x } 
               \int_{-\infty}^{\infty} e^{-y} 
\{ e^{-\frac{\pi}{2}i\gamma_1} lA_{-}(y) +  
e^{\frac{\pi}{2}i\gamma_2} l{\overline A}_{-}(y) \} dy
			     \Bigr). \np
\label{scaleLam}
\end{eqnarray}
Thanks to the subsidiary condition for the additional zero
$\theta$ (\ref{sub}), we have
\begin{eqnarray}
e^{-\overline{\theta}} &&=\pi - \frac{2i}{\pi} \biggl(
      \int_{-\infty}^{\infty} 
      F\left(\frac{2}{\pi}(z-\overline{\theta})+
             i(1-\gamma_1)\right)lA_{+}(z)dz  \np
  -&& \int_{-\infty}^{\infty}
   F\left(\frac{2}{\pi}(z-\overline{\theta})- 
      i(1-\gamma_2)\right)l\overline{A}_{+}(z)dz
 \biggr).
 \label{scaletheta}
\end{eqnarray}
%
%
For further simplification, we define $D_{\pm}$ by,
\begin{eqnarray}
D_{\pm} &:= &\int_{-\infty}^{\infty}
   \Bigl( lA_{\pm}(x) \frac{d}{dx}la_{\pm}(x)+l\overline{A}_{\pm}(x) 
          \frac{d}{dx}l\overline{a}_{\pm}(x)  \np
	& &  -la_{\pm}(x)\frac{d}{dx}lA_{\pm}(x)-
	    l\overline{a}_{\pm}(x)\frac{d}{dx}l\overline{A}_{\pm}(x)\Bigr) dx  \np
&=& \int_{a_{\pm}(-\infty)}^{a_{\pm}(\infty)} 
       \left( \frac{\ln(1+a)}{a}- \frac{\ln a}{1+a} \right) da \np
& &	+ \int_{\overline{a}_{\pm}(-\infty)}^{\overline{a}_{\pm}(\infty)} 
       \left(\frac{\ln(1+\overline{a})}{\overline{a}}-
			     \frac{\ln \overline{a}}{1+\overline{a}}\right) d\overline{a}.
\end{eqnarray}
Obviously, they are equal to special values of Roger's Dilogarithm ${\cal L}$,
\begin{eqnarray}
D_{\pm} &&= 
2 {\cal L} \left( \frac{a_{\pm}(\infty)}{1+a_{\pm}(\infty)} \right)+
2 {\cal L} \left( \frac{\overline{a}_{\pm}(\infty)}{1+\overline{a}_{\pm} (\infty)} \right) \np
&& - 2 {\cal L} \left(\frac{a_{\pm}(-\infty)}{1+a_{\pm}(-\infty)} \right)-
  2 {\cal L} \left(\frac{\overline{a}_{\pm}(-\infty)}
   {1+\overline{a}_{\pm} (-\infty)} \right), \np
{\cal L}(x)&&:=-\frac{1}{2} \int_{0}^{x} dy \left[ \frac{\ln (1-y)}{y} +
\frac{\ln y}{1 - y} \right]. 
\eea
We then apply the dilogarithm trick to (\ref{scaleNLIE1})--(\ref{scaleNLIE4}).
For example, we take the first two equations.
After differentiating, 
we multiply them by $lA_+(x),l\overline{A}_+(x)$ respectively
and take the summation. 
We call resultant equality (A).
Next multiply (\ref{scaleNLIE1}),
(\ref{scaleNLIE2}) by $(lA_+(x))',
(l\overline{A}_+(x))'$, respectively and 
take the summation.
Let us call the outcome as (B). 
Finally we subtract (B) from (A) and integrate over $x$.
The lhs of the equality is nothing but $D_+$.
Remarkably in the rhs, most complicated terms like
\begin{eqnarray}
& & - \int lA_+(x) \frac{d F_2(x-y)}{dx} l\overline{A}_+(y) dx dy \np
& & = - \int lA_+(x) F_2(x-y) \frac{d l\overline{A}_+(y)}{dy} dx dy, 
\end{eqnarray}
and 
\be
\int \frac{dl \overline{A}_+(x)}{dx} \overline{F}_2(x-y) l {A_+}(y) dx dy,
\end{equation}
cancel with each other. 
After rearrangement we obtain,
\begin{eqnarray}
& &D_+ + 2\pi i {\cal F}(\infty) \ln \frac{A_{+}(\infty)}{\overline{A}_{+}(\infty)}
\np
& & = \int_{-\infty}^{\infty} 2 e^{-y} 
  \left( e^{\frac{\pi}{2}i\gamma_1} lA_+(y) + 
  e^{- \frac{\pi}{2}i\gamma_2} l\overline{A}_+(y) \right)  dy  \np
& & + 8 i \int_{-\infty}^{\infty} F\left(\frac{2}{\pi}(x-\overline{\theta})+
      i(1-\gamma_1)\right)lA_{+}(x) dx \np
& & - 8 i \int_{-\infty}^{\infty} F\left(\frac{2}{\pi}(x-\overline{\theta})-
     i(1-\gamma_2)\right)l\overline{A}_{+}(x) dx,
\label{trick1}
\end{eqnarray}
where $a_+(-\infty)=\overline{a}_+(-\infty)=0$ is used.
Similarly, from  (\ref{scaleNLIE3}), (\ref{scaleNLIE4}), we have
\begin{eqnarray}
& & D_- + 2\pi i {\cal F}(-\infty) \ln \frac{A_{-}(\infty)}{\overline{A}_{-}(\infty)} \np
  & & =  \int_{-\infty}^{\infty} 2 e^{-x} \left( e^{- \frac{\pi}{2}i \gamma_1} lA_-(x) +
e^{\frac{\pi}{2}i\gamma_2} l\overline{A}_-(x) \right) dx, \np
\label{trick2}
\end{eqnarray}
Applying (\ref{trick1}), (\ref{trick2}), together with 
(\ref{scaletheta}) to (\ref{scaleLam}), 
\begin{eqnarray}
& & \ln\Lambda_{\rm fn}(x) \sim  \frac{\pi}{2}i+
             \frac{2\eta}{2\pi^2 \beta \sin 2 \eta} \np
& &  \times \Bigl\{  e^{\frac{\pi}{2} x} \left( - 4 \pi^2 + D_+ + 
  2\pi i {\cal F}(\infty) \ln \frac{A_{+}(\infty)}{\overline{A}_{+}(\infty)}  \right) \np
& & \ \ \ \ + e^{-\frac{\pi x}{2}} \left( D_- + 2\pi i {\cal F}(-\infty) 
  \ln \frac{A_{-}(\infty)}{\overline{A}_{-}(\infty)} \right) \Bigr \}. 
\label{trickLam}
\end{eqnarray}
Now that the asymptotic values are easily found,
\bea
{\cal F}(\infty)&&=-{\cal F}(-\infty) = 
    \frac{\pi-4\eta}{4(\pi-2\eta)}, \np
a_+(\infty)&&=\overline{a}_-(\infty)=e^{(\pi-4\eta)i}, \np
a_-(\infty)&&=\overline{a}_+(\infty)=e^{(-\pi+4\eta)i},
\eea
we can explicitly evaluate (\ref{trickLam}) 
at $x=0$,
\be
 \ln\Lambda_{\rm fn}(x=0) = 
 \frac{\pi}{6\beta v_{\rm F}}
 -\frac{\pi}{\beta v_{\rm F}} 
  \left( \frac{1}{\alpha}+\frac{\alpha}{4} \right)+\frac{\pi}{2}i. 
\end{equation}
where ${\cal L}(x)+{\cal L}(1-x)=\pi^2/6$ is also applied.
Here $\alpha$ is introduced in (\ref{alpha}) and
the Fermi velocity $v_{{\rm F}}$ is also derived
in (\ref{fv2}) for $n_{\rm e}=0.5$. 
The first term is identical to the largest eigenvalue sector, and it
reproduces conformal anomaly term with $c=1$.
Comparing them, one concludes
\be
\frac{\Lambda_2}{\Lambda_1} \sim e^{ i k_{{\rm F}} -1/\xi},
\label{eqkf}
\end{equation}
where  $ k_{{\rm F}}$ denotes the ``Fermi momentum".
Note that $k_{\rm F}=\pi/2$ in the half-filling case.
Consequently the inverse correlation length is given as
\begin{equation}
\xi^{-1} = \frac{\pi T}{v_{{\rm F}}}\left(\frac{1}{\alpha}+\frac{\alpha}{4}
\right),
\end{equation}
These are nothing but the expected results from CFT (see (\ref{xitcft})).
This fact represents the consistency of both our result and
validity of CFT mapping in the finite temperature problem at low temperatures.
%
\subsection{ Numerical Analyses on NLIE}
%
Having verified consistency at the specific limits, we now 
perform numerical analyses on the NLIE for 
a wide range of temperatures,
electron fillings and interaction strengths.

To keep the electron filling constant, we adopt the
temperature dependent
chemical potential which are determined by the curve,
\be 
\frac{d\langle n_{\rm e} (T, \mu(T))\rangle}{dT}=
\frac{d}{dT}\left(\frac{\partial f}{\partial\mu}\right)_T=0.
\label{chem}
\end{equation}
The NLIE are numerically solved by the iteration
method. 
In each iteration steps,
convolution parts are treated by the Fast Fourier 
Transformation (FFT).
As a technical remark,  we call an attention to 
proper re-scaling of auxiliary functions for the FFT;
one needs to modify the integrands such that these 
asymptotic values vanish. 
{}From the asymptotics in 
(\ref{asymptotics1}) and (\ref{asymptotics2}),
we introduce
\be
 \mfB(x) :=\begin{cases}
  \mfA(x)/\mfA(\infty) 
  & \text{for $x\ge 0$,} \\
  \mfA(x)/\mfA(-\infty)
  & \text{for $x<0$,}
  \end{cases}
\end{equation}
and similarly for others. 
We also rewrite NLIE in terms of  ${\mathfrak{B}}(x)$,
which has now zero asymptotic values.
For example,
\begin{eqnarray}
   \ln {\mathfrak{a}}&&(x)= 
     -\frac{\pi \beta  \sin(2\eta)}{4\eta \cosh\frac{\pi}{2}(x-i\gamma_1)}+
      F*\ln\mfB(x) \np
    &&-F*\ln\overline{\mfB}(x+2i-i(\gamma_1+\gamma_2))
      +{\cal F}(x) \ln \frac{\mfA(\infty)}{\mfA(-\infty)} \np
      &&-{\cal F}(x+2i-i(\gamma_1+\gamma_2)) 
       \ln \frac{\overline{\mfA}(\infty)}{\overline{\mfA}(-\infty)} \np
    &&+2\pi i {\cal F}(x-\theta+i(1-\gamma_1))  
    +\beta\mu.
\eea
In addition, one must be careful in the branch cuts of the 
logarithms.
In the above, 
$\ln \mfA(\infty)/\mfA(-\infty)$ and so on
must be understood as
\be
\ln \frac{{\mathfrak{A}}(\infty)}{{\mathfrak{A}}(-\infty)}=
\ln \left( -\frac{\sinh(\beta \mu/2-2i\eta)}{\sinh(\beta \mu/2 +2i\eta)} 
\right)
  + (\pi-4\eta) i.
\end{equation}
Under these arrangements, the iteration method works in a stable manner.

We plot the temperature dependence of the 
correlation length $\xi T$ in FIG.~\ref{cl1} for various fillings 
keeping the interaction
strength constant $\Delta= \cos (\pi/6)$.
\begin{figure}
\begin{center}
\includegraphics[width=0.48\textwidth]{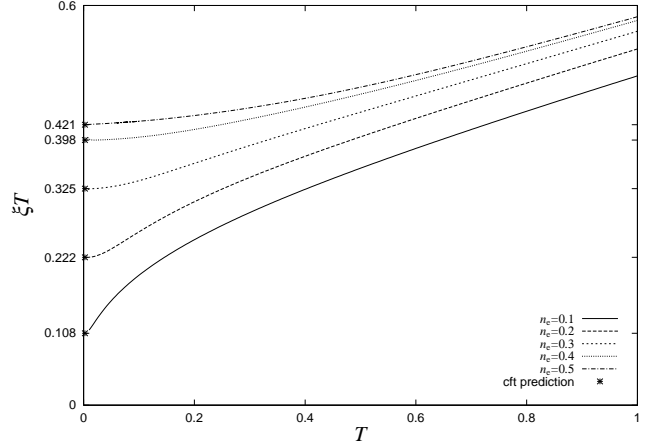}
\end{center}
\caption{The temperature dependence of
the correlation length $\xi$ for $p_0$=6.}
\label{cl1}
\end{figure}
The extrapolated values $T \rightarrow 0$ agree with the predictions from
CFT within few percents even far away from ``half-filling"
($n_{\rm e}=0.5$).
The curves are going down gradually with the decrease of electron density $n_{\rm e}$.
As further information, chemical potential $\mu(T)$ determined by
Eq.~(\ref{chem}) and
the location of the additional zero $\theta$
are depicted in FIG.~\ref{mu} and FIG.~\ref{sfmzero6}, 
respectively.
%
%
The zero $\theta$ moves on a smooth curve and its curvature 
increases with the decrease of $n_{\rm e}$. 
In fact, we find that it moves to $\theta=i$ 
when $n_{\rm e}, T  \rightarrow 0$.
(See also the analytic argument for non-interacting Fermion case
 in FIG.~\ref{fzeros} for $\mu=1.0$.)
We also calculate the ``Fermi momentum"  
$k_{\rm{F}}=\Im \ln \Lambda_2/\Lambda_1$
(cf. Eq.~(\ref{eqkf})).
(Here the inverse period of oscillatory behavior at
arbitrary $T$ is referred to as $k_{\rm F}$ as in the case of $T=0$.)
The figure clearly shows the temperature dependency of $k_{\rm F}$.
In the low temperature limit 
$T \rightarrow$ 0, it converge to the expected value, 
$k_{{\rm F}} = n_{\rm e} \pi$,
which indicates the significance of the Fermi surface for one-particle
excitations in the Luttinger liquid  at $T=0$.\cite{KY}
With the increase of $T$, the auxiliary functions cease to
exhibit a sharp crossover behavior (\ref{crossover}), 
which roughly corresponds 
to broadening of the Fermi distribution at $T>0$.
The particle excitations are enhanced within the wide range 
near the Fermi surface, which yield the shift of $k_{\rm F}$.
We remark that such $T$ dependent
oscillatory behavior has been  reported for the longitudinal
correlation function of ferromagnetic  Heisenberg model. \cite{FKM2}
Although the physical origins are  different for these two cases,
the explicit determination of $T$ dependency is important.
\begin{figure}
\begin{center}
\includegraphics[width=0.48\textwidth]{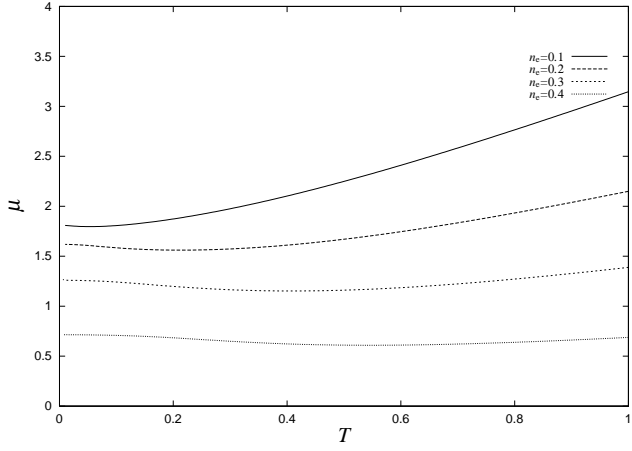}
\end{center}
\caption{The temperature dependence of
the chemical potential $\mu$ for $p_0$=6.}
\label{mu}
\end{figure}
\begin{figure}
\begin{center}
\includegraphics[width=0.48\textwidth]{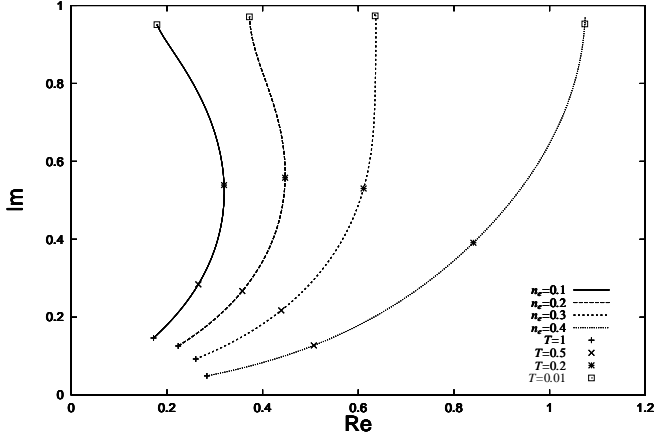}
\end{center}
\caption{The trajectory of the 
additional zero $\theta$ inside the physical strip.}
\label{sfmzero6}
\end{figure}
\begin{figure}
\begin{center}
\includegraphics[width=0.48\textwidth]{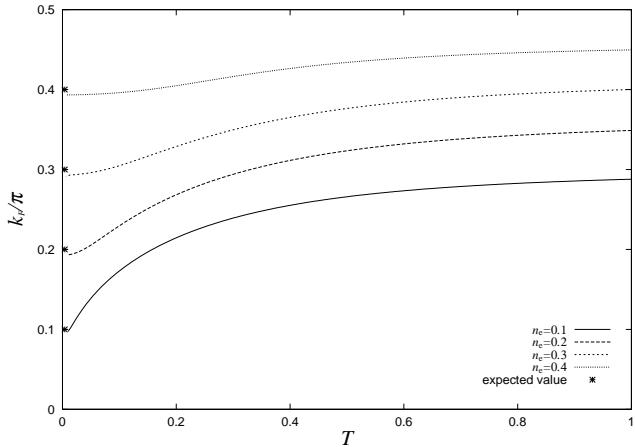}
\end{center}
\caption{The temperature dependence of
the Fermi momentum $k_{{\rm F}}$ for $p_0$=6.}
\label{kf}
\end{figure}
FIG.~\ref{sfmcl01}, \ref{sfmcl04}
presents the temperature dependence of the correlation length
for various interaction strengths
for fixed $n_{\rm e}$.
Naturally in the limit, $n_{\rm e}, T\rightarrow 0$, 
$\xi T$ does not depend significantly on
the interaction strength; 
it merely behaves as 
$\xi T \sim v_{{\rm F}}/\pi \sim n_{\rm e}$
(see Appendix B).
This fact is typical for non-interacting cases.
Although our model inherits strong correlations, 
FIG.~\ref{sfmcl01} indicates that $n_{\rm e}=0.1$ is already well 
described by  ``non-interacting approximation" 
and also shows this approximation is applicable
in the wide range of $T$.
On the other hand, data for $n_{\rm e}=0.4$ show strong dependency on
$\Delta$, therefore it belongs to proper ``interacting class"
(see FIG.~\ref{sfmcl04}).
It seems these crossover occurs near $n_{\rm e} \sim 0.25 $ but
it is not yet conclusive.
We hope to clarify this in a future communication.
\begin{figure}
\begin{center}
\includegraphics[width=0.48\textwidth]{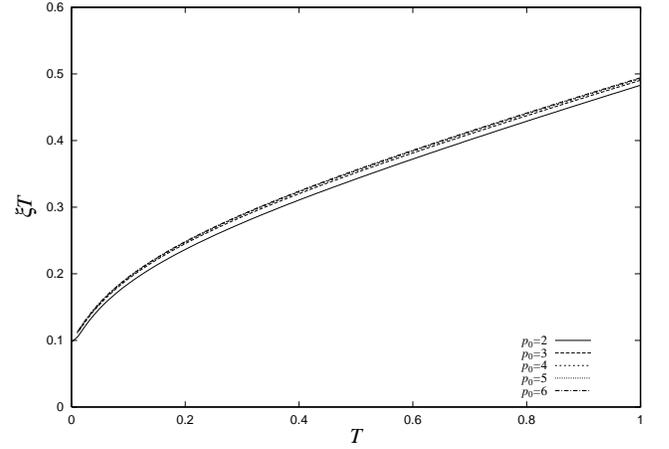}
\end{center}
\caption{The temperature dependence of
the correlation length  for $n_{\rm e}=0.1$.}
\label{sfmcl01}
\end{figure}
\begin{figure}
\begin{center}
\includegraphics[width=0.48\textwidth]{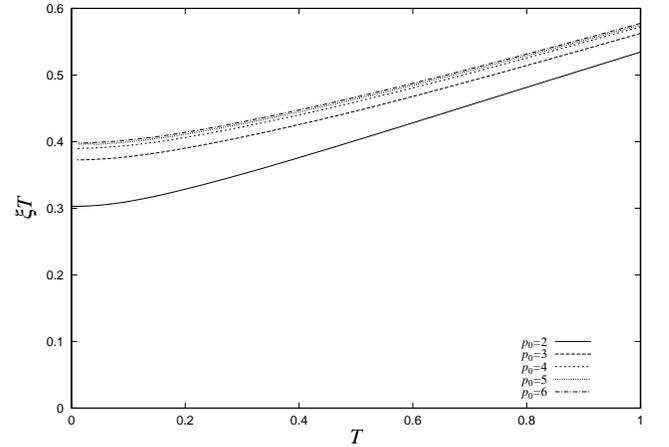}
\end{center}
\caption{The temperature dependence of
the correlation length  for $n_{\rm e}$=0.4.}
\label{sfmcl04}
\end{figure}

Finally we plot the correlation length of 
transverse spin-spin correlation 
$\langle\sigma^{+}_j\sigma_j^{-}\rangle$ without
external field (FIG.~\ref{clxxz}) for comparison with
$n_{\rm e}=0.5$ of spinless Fermion models
(FIG.~\ref{sfmcl}).
Besides the difference between their limiting values
at $T \rightarrow 0$, one clearly sees the difference in
the dependence of $\xi T$ on $T$.
\begin{figure}
\begin{center}
\includegraphics[width=0.48\textwidth]{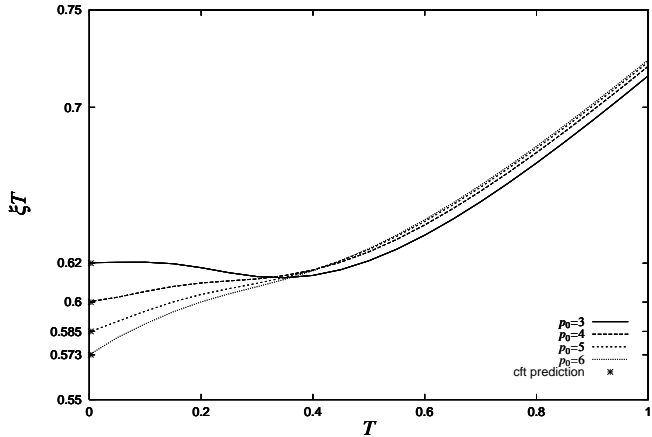}
\end{center}
\caption{The correlation length for $\langle
\sigma^{+}_j\sigma^{-}_i\rangle$ of the
corresponding ${XXZ}$ model with zero magnetic field.}
\label{clxxz}
\end{figure}
\begin{figure}
\begin{center}
\includegraphics[width=0.48\textwidth]{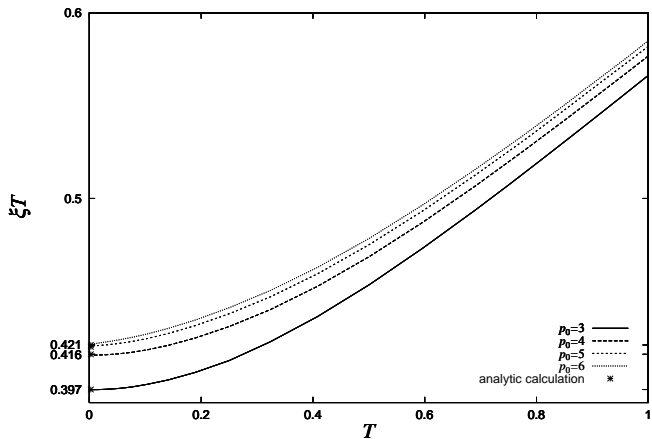}
\end{center}
\caption{The temperature dependence of
the correlation length at half-filling.}
\label{sfmcl}
\end{figure}
%
\section{Summary and discussion}
We have proposed the QTM approach to the integrable 
lattice Fermion systems at any finite temperatures.
The Fermionic $R$-operator,
together with its super-transposition
$\widetilde{R}$, where
the Fermion statistics is embedded naturally,
play the crucial role in
this approach.
Consequently, we have observed the
significant difference between the  
Fermion model and
that of the spin model.
In principle, 
we can apply this approach to
any integrable 1D Fermion systems.
The application to
the Hubbard model is under progress.

Here we comment on the ``attractive regime"
$t>0$, $\Delta<0$ in (\ref{eq.hamiltonian}),
which we have not been concerned with in this
paper.
In the ${XXZ}$ model
without external magnetic field,
one may recall the remarkable difference 
between the repulsive (anti-ferromagnetic) 
case and the attractive (ferromagnetic) one.
\cite{FKM,FKM2}
In the repulsive regime,
the eigenvalues related
to the correlation  $\langle
\sigma^{+}_j\sigma^{-}_i\rangle$ or
$\langle\sigma^{z}_j\sigma^{z}_i\rangle$
is characterized by  two real 
additional zeros 
which are symmetric with respect to 
the imaginary axis.
This symmetry is never broken at any
temperatures.
On the other hand in the
attractive regime, ``level crossing"
occurs successively.
One may attribute it to
the change of the distribution patterns
of the additional zeros.
It will be interesting to see
if similar phenomena happens for the 
spinless Fermion model in the attractive
regime.

Finally we refer to another formulation of NLIE 
derived from the different choice of the auxiliary 
functions.
The NLIE have a close connection
with the ``TBA" or ``excited states TBA"
equations from the standard ``string 
hypothesis".

The idea is as follows.
First we embed the QTM itself
into a more general family called $T$-functions
and explore functional relations among them 
($T$-system).
Then we define the $Y$-functions by
certain ratio of the $T$-functions
and also derive functional relations 
for them ($Y$-system).
The analytical properties of these functions
leads to the NLIE which determine the
free energy and the correlation length.
As concerns the largest eigenvalue sector,
the $T$-functions coincide with those
in Ref.~\onlinecite{KSS}.
Therefore the derived NLIE for the free energy
are identical to 
the TBA equations
of the ${XXZ}$ model.\cite{TS,KSS}
In contrast, 
for the second largest eigenvalue sector
we find the essential difference between
the Fermion model and the corresponding
spin model.
For example, we write explicitly
the NLIE (excited state TBA equation)
for $p_0=5$, $\mu=0$  as 
\bea
\ln\eta_1(x)&=&-\frac{5\beta\sin\frac{\pi}{5}x}
   {2\cosh\frac{\pi}{2} x}+K*\ln(1+\eta_2)(x)+\pi i,\nn \\
\ln\eta_2(x)&=&K*\ln(1-\eta_1)(1-\eta_3)(x) \nn \\
&+&\ln\left(\tanh\frac{\pi}{4}(x-\theta_1)
         \tanh\frac{\pi}{4}(x-\theta_2)  
    \right)+\pi i, \nn \\
\ln\eta_3(x)&=&K*\ln(1+\eta_2)(1-\kappa^{2})(x) \nn \\
\ln\kappa&=&K*\ln(1-\eta_2)(x)+
    \ln\left(\tanh\frac{\pi}{4}(x-\theta_2)\right) \nn \\
&+&\frac{\pi}{2} i,
\eea
where $\theta_1$ and $\theta_2$ are
determined from
\bea
&&i\frac{5\beta\sin\frac{\pi}{5}\theta_1}
   {2\sinh\frac{\pi}{2} \theta_1}+
    K*\ln(1+\eta_2)(\theta_1+i)-\pi i=0, \nn \\
&&K*\ln(1+\eta_2)(1-\kappa^{2})(\theta_2+i)=0.
\eea
The meaning of the functions $\eta_j$ and the quantities
$\theta_j$ are similar to those in Ref.~\onlinecite{KSS}.
Although the above expressions 
are quite different from those in Sec.~III,
the numerical result shows a good agreement.
The detailed derivations of above equations 
will be described in a
separate communication.\cite{S} 
\acknowledgments

The authors are grateful to A. Kuniba, M. Wadati 
for helpful comments and continuous encouragements.
M. S. thanks H. Asakawa for discussions.
J. S. thanks 
A. Kl{\"u}mper, R. Martinez, B. M. McCoy and C. Scheeren
for useful discussions.
This work is in part supported by Grant-in-Aid for JSPS Fellows from
the Ministry of Education, Science, Sports and Culture of Japan.
\appendix
\section{Diagonalization of the Quantum Transfer Matrix}
%
Here we shall diagonalize the  QTM (\ref{eq.QTM})
by means of the algebraic Bethe ansatz.

First let us recall that the monodromy operator (\ref{eq.QTMmonodromy}) 
satisfies the global Yang-Baxter relation
\begin{eqnarray}
& & {\cal R}_{21}(v-v') {\cal T}_{1}(u_N,v) {\cal T}_{2}(u_N,v') \nonumber \\
& & = {\cal T}_{2}(u_N,v') {\cal T}_{1}(u_N,v) {\cal R}_{21}(v-v').
\label{eq.QTMYBE2}
\end{eqnarray}
Writing the monodromy operator as
\begin{eqnarray}
{\cal T}_{j}(u_N,v) & = &  A(v) \ (1 - n_j)  + B(v) \ c_j \nonumber \\
& & + C(v) \  c_j^{\dagger} + D(v) \ n_j, \ \ \ \ j=1,2, 
\end{eqnarray}
and substituting them into (\ref{eq.QTMYBE2}),
we get the commutation relations among the operators ${A(v), \cdots, D(v)}$,
\begin{subequations}
\begin{eqnarray}
A(v) B(v')  &=& \frac{a(v'-v)}{b(v'-v)} B(v') A(v) \nonumber \\
& & - \frac{c(v'-v)}{b(v'-v)} B(v) A(v'), \label{eq.commA} \\
D(v) B(v')  &=& - \frac{a(v -v')}{b(v-v')} B(v') D(v) \nonumber \\
& &  + \frac{c(v-v')}{b(v-v')} B(v) D(v'), \label{eq.commB} \\
B(v) B(v') &=& B(v') B(v). \label{eq.commC}
\end{eqnarray}  
\end{subequations}
To derive these relations, one should pay attention to the fact that 
${B(v)}$ and ${C(v)}$  anti-commute with 
the Fermion operators ${c_j}$ and ${c_j^{\dagger}}$. 

The commutation relations (\ref{eq.commA})--(\ref{eq.commC}) are 
quite similar to the corresponding ones for the ${XXZ}$ model.\cite{KBI} 
In fact the relations (\ref{eq.commA}) and (\ref{eq.commC}) are identical. 
The second relation (\ref{eq.commB}), however, is different: 
there appears an overall ``minus" sign on the rhs. 

Now we define the reference state by  
\begin{eqnarray*}
| \Omega \rangle := \prod_{m=1}^{N/2} |0 \rangle_{a_{2m}} 
 \sotimes | 1 \rangle_{a_{2m-1}},
\end{eqnarray*} 
\begin{equation}
| 1 \rangle_{a_{2m-1}} := c_{a_{2m-1}}^{\dagger} |0 \rangle_{a_{2m-1}}.
\end{equation} 
Then using the relations,
\begin{eqnarray}
\widetilde{R}_{a_{2m-1},j}&&(v-u_N) | 1 \rangle_{a_{2m-1}} \np 
&&=- a(-v + u_N) n_j \ | 1 \rangle_{a_{2m-1}} \np 
&&+b(-v + u_N) (1-n_j) \ | 1 \rangle_{a_{2m-1}} 
+c_{j} | 0 \rangle_{a_{2m-1}}, \np
\end{eqnarray}
and
\begin{eqnarray}
 R_{a_{2m},j}(v + u_N) | 0 \rangle_{a_{2m}} &=&
a(v + u_N) (1- n_j) \ | 0 \rangle_{a_{2m}} \nonumber \\ 
&+&  b(v + u_N) n_j \ | 0 \rangle_{a_{2m}} - c_{j} | 1 \rangle_{a_{2m}}, \nonumber \\
\end{eqnarray}
we find that
\begin{eqnarray}
A(v) | \Omega \rangle &=& \left(a(v + u_N) b(-v + u_N)\right)^{N/2} | 
       \Omega \rangle, \\
D(v) | \Omega \rangle &=& \left( - b(v + u_N) a(-v + u_N) \right)^{N/2} | 
       \Omega \rangle. 
\end{eqnarray}
Hence the state ${| \Omega \rangle}$ is an eigenstate of the QTM (\ref{eq.QTM}) 
with the eigenvalue
\begin{eqnarray}
\Lambda_{0}(v) &=& \left( \frac{ \sin \eta(v + u_N + 2) \sin \eta (-v + u_N)}{\sin^2 2 \eta} \right)^{N/2} \nonumber \\
& & + \left( - \frac{ \sin \eta (v+u_N) \sin \eta ( -v + u_N + 2)}{\sin^2 2 \eta} \right)^{N/2}. \nonumber \\
\end{eqnarray} 
An eigenstate with $N_{{\rm e}}$ 
``particles" can be constructed by multiplying the operators ${B(v_j)}$ 
to the reference state 
\begin{equation}
| \Psi \rangle := 
\prod_{j=1}^{N_{\rm e}} B(v_j) | \Omega \rangle. \label{eq.state}
\end{equation}
Indeed, using the standard argument of the algebraic Bethe ansatz,\cite{KBI} 
we can show that the state (\ref{eq.state}) 
becomes the eigenstate of the QTM if the spectral parameters 
${v_j}$ fulfill the Bethe ansatz equations
\begin{eqnarray}
& & \left[ \frac{\sin \eta (-v_j + u_N) \sin \eta (v_j + u_N + 2)}
{\sin \eta (v_j + u_N) \sin \eta (-v_j + u_N + 2)} \right]^{N/2} \nonumber \\
& & = -  (-1)^{N/2 + N_{\rm e}}  \prod_{k =1}^{N_{\rm e}}
\frac{\sin \eta (v_j - v_k + 2)}{\sin \eta (v_j - v_k - 2)}.
\label{eq.bae2}
\end{eqnarray}
The corresponding eigenvalue of the QTM (\ref{eq.QTM})
\begin{equation}
 T_{\rm QTM}(u_N,v)| \Psi \rangle = \Lambda(v) | \Psi \rangle
\end{equation}
is given by
\begin{eqnarray}
& & \Lambda(v) =  \left( \frac{\sin \eta (v + u_N + 2)  
           \sin \eta (-v + u_N)}{\sin^2 2 \eta }   \right)^{N/2} \nonumber \\
& & \hspace{1cm} \times \prod_{j=1}^{N_{\rm e}} 
          \frac{ \sin \eta (v - v_j - 2)}{ \sin \eta (v - v_j)} \nonumber \\
& & + (-1)^{N/2 + N_{\rm e}} \left( \frac{\sin \eta (v + u_N) 
           \sin \eta  (-v + u_N + 2)}{\sin^2 2 \eta} \right)^{N/2}  \nonumber \\
& & \hspace{1cm} \times \prod_{j=1}^{N_{\rm e}}  
           \frac{ \sin \eta (v - v_j + 2)}{ \sin \eta (v - v_j)}. 
\end{eqnarray}
%
\section{$T\ll 1$ behavior and prediction from CFT}
%
We summarize the known results of the correlation function 
at $T=0$ 
and its $T\ll 1$ behavior predicted from CFT.\cite{KBI,KY}

Let us start with the zero temperature case.
One-particle Green's 
function shows an oscillatory behavior due to the Fermi surface,\cite{KY}
\begin{equation}
\langle c^{\dagger} (x) c(0) \rangle \sim \cos(k_{{\rm F}} x) /x^{2\triangle} 
\label{correlation}
\end{equation}
The scaling dimension $\triangle$ is evaluated from 
the energy spectra in the finite size system,
\begin{equation}
\triangle = \frac{1}{4 Z({\cal K}_{\rm F})^2} (\triangle N)^2+ 
Z({\cal K}_{\rm F})^2 (\triangle D)^2 .
\label{triangle}
\end{equation}
Here $Z({\cal K}_{\rm F})$ 
is the dressed charge and ${\cal K}_{\rm F}$ denotes the ``Fermi surface" 
satisfying
\begin{eqnarray}
 Z(x)&+& \frac{1}{2\pi} \int_{-{\cal K}_{\rm F}}^{{\cal K}_{\rm F}} 
 R(x-y) Z(y) dy =1, \np
 R(x) &=& \frac{2 \sin 4 \eta }{\cosh 2x - \cos 4\eta}. 
\end{eqnarray}
$\triangle D$ and $\triangle N$ are 
(half-)integers constrained by a selection rule,
$\triangle D =\triangle N/2 $ mod 1. 
For one-particle Green's function, they are given by
$\triangle D=1/2$ and $\triangle N=1$.
Thus the critical exponent $\eta_{\rm F}$ is defined as
\be
\eta_{{\rm F}} := 2 \triangle = \frac{1}{2}\left(Z({\cal K}_{\rm F})^2 + 
\frac{1}{Z({\cal K}_{\rm F})^2} \right). 
\end{equation}
The dressed charge $Z({\cal K}_{\rm F})$ is explicitly
evaluated for two special cases; \cite{KBI}
\be
Z({\cal K}_{\rm F})=
\begin{cases}
1 &\text{for $n_{\rm e}=0$ (${\cal K}_{\rm F}=0$),} \\
\sqrt{\alpha/2}
  &\text{for $n_{\rm e}=0.5$ (${\cal K}_{\rm F}=\infty$),} 
\end{cases}
\end{equation}
where $\alpha$ is  
\be
\alpha=\frac{\pi}{\pi-2\eta}.
\label{alpha}
\end{equation}
Then the critical exponent $\eta_{\rm F}$ is given by
\be
\eta_{\rm F}=
\begin{cases}
1 &\text{for $n_{\rm e}=0$,} \\
1/\alpha+\alpha/4
  &\text{for $n_{\rm e}=0.5$.} 
\end{cases}
\end{equation}

In the scaling limit where CFT is valid, 
the correlation functions at $T\ll 1$ 
are recovered 
by the replacement 
\be
x\to \frac{v_{\rm F}}{\pi T} \sinh \frac{\pi T x}{v_{\rm F}}.
\end{equation} 
in the denominator in (\ref{correlation}). 
Here $v_{\rm F}$ denotes the Fermi velocity
\be
v_{\rm F} :=\frac{1}{2\pi\rho(x)}
            \frac{\partial \varepsilon(x)}{\partial x}
            \biggr|_{x=\cal{K}_{\rm F}}.
\label{fv}
\end{equation}
Note that $\rho(x)$ and $\varepsilon(x)$
are the density function and the dressed energy 
defined by
\bea
\rho(x)+\frac{1}{2\pi} \int_{-{\cal K}_{\rm F}}^{{\cal K}_{\rm F}} 
R(x-&&y)\rho(y)dy=\frac{\sin2\eta}{\pi(\cosh2x-\cos2\eta)}, \np  
\varepsilon(x)+\frac{1}{2\pi} \int_{-{\cal K}_{\rm F}}^{{\cal K}_{\rm F}}  
R(x-&&y)\varepsilon(y) dy \np
&&=-\frac{\sin^2 2 \eta}{\cosh 2x - \cos 2\eta} +\mu.
\eea
Thus the long distance behavior of one-particle Green's function is given by
\begin{equation}
\langle c^{\dagger} (x) c(0)\rangle
 \sim \cos(k_{{\rm F}} x) x^{-\pi \eta_{{\rm F}} |x| T/v_{{\rm F}}}. 
\end{equation}
Consequently the correlation length at $T\ll 1$ is identified with
\begin{equation}
\xi = \frac{v_{{\rm F}}}{\pi \eta_{{\rm F}} T}.
\label{fermionxit}
\end{equation}
The Fermi velocity (\ref{fv}) is analytically
calculated for the cases $n_{\rm e}\ll 1$ and $n_{\rm e}=0.5$
\be
v_{\rm F}=
   \begin{cases}
     \sim \pi n_{\rm e} &\text{for $n_{\rm e}\ll 1$,}\\
     \pi\sin2\eta/4\eta &\text{for $n_{\rm e}=0.5$.}
   \end{cases}  
\label{fv2}
\end{equation}
Therefore we get the explicit correlation length (\ref{fermionxit})
for these two special cases.
\be
\xi T=
   \begin{cases}
     \sim n_{\rm e} &\text{for $n_{\rm e}\ll 1$,}\\
     \sin2\eta/(4\eta(1/\alpha+\alpha/4)) &\text{for $n_{\rm e}=0.5$.}
   \end{cases}
   \label{xitcft}
\end{equation}
We have also verified the extrapolations from the NLIE  
agree with the prediction (\ref{fermionxit}).

Finally we remark on the spin correlation.
The main contribution to
the transverse correlation function 
simply decays algebraically,
\be
\langle \sigma^+ (x) \sigma(0) \rangle
\sim 1/x^{2\triangle'}, 
\end{equation}
that has no oscillation term.
Here $\triangle'$ takes the {\it identical} form (\ref{triangle}).
However we have to use $\triangle N=1$ and $\triangle D=0$ this time.
The difference in selection rules for these integers, which originates from
the difference in statistics, leads  to a conclusion 
\be
\triangle \ne \triangle'=\frac{1}{4Z({\cal K}_{\rm F})^2}
\end{equation}
The corresponding correlation length is given by (\ref{fermionxit}),
replacing $\eta_{{\rm F}}$ by  $\eta_{\rm S} = 1/2(Z({\cal K}_{\rm F}))^2$.
One thus obtain different correlation lengths simply according to 
the selection rules.
%
\section{derivation of NLIE}
For simplicity in notation we define 
\begin{eqnarray}
{\mathfrak{c}}(x) &:=& {\mathfrak{a}}(x+i\gamma_1),\qquad 
{\mathfrak{C}}(x):= 1+{\mathfrak{c}}(x), \np
\overline{{\mathfrak{c}}}(x) &:=& {\mathfrak{a}}(x-i\gamma_2), \qquad 
\overline{{\mathfrak{C}}}(x):= 1+\overline{{\mathfrak{c}}}(x). 
\label{cdef}
\end{eqnarray}
That is, we forget additional shifts for a moment.

We identify the analytic strips,
\begin{eqnarray}
Q(x) &:&     \qquad \Im x \in (-2p_0,0)  \np
\phi_-(x)&:& \qquad  \Im x \in [0, 2p_0)   \np
\phi_+(x)&:& \qquad  \Im x \in (-2p_0, 0]. 
\end{eqnarray}

The following identities are direct consequence of the definitions,
\begin{eqnarray}
    \Lambda(x+i)&&={\mathfrak{C}}(x)\frac{Q(x-i)}{Q(x-(2p_0-1)i)}  \np
      & & \times \phi_-(x+i) \phi_+(x-i(2p_0-3))e^{-\beta \mu/2}
               \np
\Lambda(x-i)&&=(-1)^{N/2+N_{{\rm e}}}
          \overline{{\mathfrak{C}}}(x) \frac{Q(x-(2p_0-1)i)}{Q(x-i)} \np
 && \times \phi_+(x-i) \phi_-(x+i(2p_0-3))e^{\beta \mu/2}.
\label{CTrel}
\end{eqnarray}

Now we consider the second largest eigenvalue
case $N_{{\rm e}}=N/2-1$.
We are in position to utilize the
knowledge of zeros of $\Lambda_2(x)$.

Consider the integral,
$$
\int_{\cal C} \frac{d}{d z} \ln \Lambda_2(z) e^{i k z} dz,
$$
where ${\cal C}$ encircles the edges of ``square" :
$[z_1, z_2]\cup[z_2,z_3]\cup
[z_3,z_4]\cup[z_4, z_1]$
in the counterclockwise manner,
where $z_1=-\infty-i$,
$z_2=\infty-i$,
$z_3=\infty+i$,
$z_4=-\infty+i$.
There is one zero of $\Lambda_2(x)$ in the region inside ${\cal C}$.
Thus Cauchy's theorem is applied,
\begin{eqnarray}
2 \pi i e^{i k\theta}&= &
\int^{\infty}_{-\infty}  \frac{d}{d x} \ln \Lambda_2(x-i) e^{ik (x-i)} dx \np
 & &-\int^{\infty}_{-\infty} \frac{d}{d x} \ln \Lambda_2(x+i) e^{ik (x+i)} dx.
\end{eqnarray}
One substitutes Eq.~(\ref{CTrel}) 
into the above equation and derives  identities among
the Fourier components of  logarithmic derivatives of
$Q$, ${\mathfrak{C}}$ and $\overline{\mathfrak{C}}$.
Explicitly we have,
\begin{eqnarray}
\dl Q[k]=&&-\frac{e^{k(p_0-1)}\dl{\mathfrak{C}}[k]-
                  e^{k(p_0+1)}\dl\overline{\mathfrak{C}}[k]}
                 {4\sinh(p_0-1)k\cosh k}   \nn \\
        &&+\frac{e^{k(2p_0-1)}\dl\phi_-[k]+e^k\dl\phi_+[k]}
               {2\cosh k} \nn \\
        &&-\frac{\pi i e^{k(p_0+i\theta)}}{2\sinh(p_0-1)k\cosh k}.
\label{Qsol}
\end{eqnarray}
In the above we adopt a notation 
\be
\widehat{dl}{\mathfrak{C}}[k] := \int_{-\infty}^{\infty} 
 \frac{d \ln {\mathfrak{C}}(x)}{dx} e^{ikx} dx,
\end{equation}
etc as the Fourier component of the logarithmic derivatives.

On the other hand, from  the definition (\ref{cdef}), 
we have
\begin{eqnarray}
\dl\mfc[k]=&&2\dl\phi_-[k]e^{kp_0}\sinh(p_0-1)k \np
         &&-2\dl\phi_+[k]e^{-(2p_0-2)k}\sinh k \np
         &&-2\dl Q[k]e^{-(p_0-1)k}\sinh(p_0-2)k \label{fourierc}
\end{eqnarray}
and similarly for $ \widehat{dl}\overline{{\mathfrak{c}}}[k]$.

One substitutes (\ref{Qsol}) into (\ref{fourierc}) to obtain a closed
equation among the Fourier modes of the auxiliary functions.

Using the explicit form for $\phi_{\pm}$,
we get,
\begin{eqnarray}
\widehat{dl}{\mathfrak{c}}[k]=&&  
    -2\pi i \frac{N}{2} \frac{\sinh u_N k}{\cosh k} 
+\widehat{dl}{\mathfrak{C}} \frac{\sinh(p_0-2)k}
   {2 \cosh k \sinh(p_0-1)k}                         \np
&&-\widehat{dl}\overline{{\mathfrak{C}}}
  \frac{e^{2k} \sinh(p_0-2)k}{2 \cosh k \sinh(p_0-1)k} \np
&&+2 \pi i e^{k+ik\theta}\frac{\sinh(p_0-2) k}{2 \cosh k \sinh(p_0-1)k}.
\label{nliec}
\end{eqnarray}
By the inverse transformation and integration over $x$ we arrive at 
NLIE. 
Note that the integration constant is determined by
the asymptotic values in (\ref{asymptotics1}) and
(\ref{asymptotics2}). 

After introducing the shifts $\gamma_{1,2}$, one obtains 
the identical NLIE in the main text,
except for ``driving terms" as we have not yet taken 
the Trotter limit $N \rightarrow \infty$.
To be precise the driving term for $\ln \mfa(x)$ is
\be
\frac{N}{2} \int_{-\infty}^{\infty}
 \frac{\sinh u_N k}{k \cosh k} e^{ik(x-i\gamma_1)}dk.
\end{equation}
Due to the combination of $u_N=-\beta\sin 2\eta/2\eta N $ 
and $N$ entering above,
the Trotter limit is carried out analytically.
Then one ends up with (\ref{NLIE}).

The expression for the eigenvalue is derived in a similar way.
One first notes the ``inversion identity",
\be
\widetilde{\Lambda}_2(x+i )\widetilde{\Lambda}_2(x-i) 
=-\psi(x)\mfC(x) \overline{\mfC}(x),
\label{inversion}
\end{equation}
where
\be
\psi(x):=\frac{\phi_+(x-i)\phi_-(x+i)}{\phi_+(x+i)\phi_-(x-i)},
\label{psi}
\end{equation}
and
\be
\widetilde{\Lambda}_2(x)=\frac{\Lambda_2(x)}{\tanh\frac{\pi}{4}(x-\theta)
                                   \phi_+(x+2i)\phi_-(x-2i)},
\end{equation}
is introduced to exclude the zeros of $\Lambda_2(x)$ and
to compensate the divergence of
$\Lambda_2(x)$ at $x\to\pm\infty$.

Then the lhs
is ANZC in a strip $\Im x\in[-1,1]$ and also
rhs is ANZC in a narrow strip including the real axis.
One thus can solve (\ref{inversion}) and we get the
expression 
\begin{equation}
\ln \Lambda_2(x)= \ln \Lambda_{\rm{gs}}(x)+ \ln \Lambda_{\rm{fn}}(x),
\label{second}
\end{equation}
where
\begin{subequations}
\begin{eqnarray}
\ln \Lambda_{\rm{gs}}(x)&:=&-\frac{N}{2}\int_{-\infty}^{\infty}
               \frac{\sinh k u_N \sinh (p_0-1)k}
                    {k\cosh k\sinh p_0 k} e^{-i k x}dk \np
             &&+\ln\phi_+(x+2i)\phi_-(x-2i), \\
\ln \Lambda_{\rm{fn}}(x)&:=&K*\ln\mfA(x+i\gamma_1)       
    +K*\ln\overline{\mfA}(x-i\gamma_2) \np 
   &&+\ln\tanh\frac{\pi}{4}(x-\theta)-\frac{\pi i}{2}. \label{finiteT}
\end{eqnarray}
\end{subequations}
Taking the Trotter limit $N\to\infty$
after setting $x=0$ and using the identity
\be
\lim_{N\to\infty}\ln\phi_+(2i)\phi_-(-2i)=-\frac{\beta}{2}\Delta,
\end{equation}
we derive the first excited free energy as 
(\ref{sec}).

Next we consider the largest eigenvalue sector $N_{\rm{e}}=N/2$.
In this case, the spinless Fermion model shares same
equations with the ${XXZ}$ model. 
Then the following NLIE  
have been already derived in Ref.~\onlinecite{KZeit}.
\begin{eqnarray}
   \ln {\mathfrak{a}}_0&&(x) =
     -\frac{\pi \beta  \sin(2\eta)}{4\eta \cosh\frac{\pi}{2}(x-i\gamma_1)}
       +F* \ln {\mathfrak{A}}_0(x) \np
      -&&F* \ln \overline{{\mathfrak{A}}}_0(x+2i-i(\gamma_1+\gamma_2))+
      \frac{\beta \mu p_0}{2(p_0-1)}, \np
\ln \overline{{\mathfrak{a}}}_0&&(x) =
     -\frac{\pi \beta  \sin(2\eta) }{4\eta \cosh\frac{\pi}{2}(x+i\gamma_2)}+
      F*\ln \overline{{\mathfrak{A}}}_0(x) \np
   - &&F*\ln {\mathfrak{A}}_0(x-2i+i(\gamma_1+\gamma_2))-
       \frac{\beta\mu p_0}{2(p_0-1)},  
\label{NLIElg}
\end{eqnarray}
where auxiliary functions $\mfa_0$ etc are 
defined by similar way to Eq.~(\ref{bdef}).
Note that their asymptotic values $|x|\to\infty$ 
are explicitly written as
\be
\mfa_0(x)= \exp(\beta\mu),\qquad  
\overline{\mfa}_0(x)=\exp(-\beta\mu). 
\label{asymptotics3}
\end{equation}

Through above NLIE, $\Lambda_1(x)$ is
described as
\bea
\ln \Lambda_1(x)&=&\ln \Lambda_{\rm{gs}}(x)+
               K*\ln\mfA_0(x+i\gamma_1) \np
              &&+K*\ln\overline{\mfA}_0(x-i\gamma_2). 
\eea
Taking the Trotter limit $N\to\infty$, 
we get the free energy per site $f$ as,
\begin{eqnarray}
f&&=-\frac{1}{\beta} \ln \Lambda_1 (0)-\frac{1}{4}{\Delta} \np 
 &&=\epsilon_0 -\frac{1}{\beta} K* \ln{\mathfrak{A}}_0(i\gamma_1) -
 \frac{1}{\beta} K* \ln \overline{{\mathfrak{A}}}_0(-i\gamma_2),\np
\label{eigenvallg}
\end{eqnarray}
where 
$\epsilon_0$ is the well-known ground state energy per site,
\be
\epsilon_0=
-\int_{-\infty}^{\infty} R(x) 
\frac{\sin^{2}2\eta}{\cosh 2\eta x -\cos 2\eta}+\frac{1}{4}\Delta.
\label{ground}
\end{equation}
Though we do not analyze (\ref{NLIElg}), (\ref{eigenvallg})
here, they are implicitly used in the evaluation of the correlation
length.
%
\section{Free Fermion model}
Here we consider the free energy and the correlation
length for the free Fermion model
$\Delta=0$ ($2\eta=\pi/2$) in (\ref{eq.hamiltonian}). 
In this case we have 
$\phi_{\pm}(x \pm 4i) = (-1)^{\frac{N}{2}}\phi_{\pm}(x)$ and 
$Q(x \pm 4i) = (-1)^{N_{{\rm e}}}Q(x)$ from (\ref{qdef}). 
Then (\ref{eq.eigen}) simplifies to
\be
\Lambda(x)=\varrho(x)\frac{Q(x+2i)}{Q(x)},
\label{freeT}
\end{equation}
where
\bea
\varrho(x)&:=&\phi_-(x-2i)\phi_+(x) e^{\frac{1}{2}\beta \mu} \nn \\
            &&+(-1)^{\frac{N}{2}}\phi_+(x+2i))\phi_-(x)
             e^{-\frac{1}{2}\beta \mu}. \nn \\
\label{defrho} 
\eea
We can easily show that
\be 
\Lambda(x+i)\Lambda(x-i)=(-1)^{N_{{\rm e}}}\varrho(x+i)\varrho(x-i).
\end{equation}
This rhs is a known function, which is a distinct feature of 
the free Fermion model.
It is convenient to modify the function $\Lambda(x)$ as
\be
\widetilde{\Lambda}(x)= \frac{\Lambda(x)}{\phi_+(x+2i)
                                  \phi_-(x-2i)},
\label{modlam} 
\end{equation}
satisfying
\bea
\widetilde{\Lambda}&&(x+i)
\widetilde{\Lambda}(x-i) \nn \\
&&= (-1)^{\frac{N}{2}+N_{{\rm e}}}\left(\psi(x)+
     \psi(x)^{-1}+2\cosh(\beta\mu)\right),
\label{eqn:rtfree}
\eea
where $\psi(x)$ has been already
defined in (\ref{psi}).

First we consider the free energy characterized
by the largest eigenvalue $\Lambda_1(x)$.
It lies in the sector $N_{{\rm e}}=N/2$.
The Bethe ansatz root $\{x^{(1)}_j\}_{j=1}^{j=N/2}$,
$\Im x^{(1)}_j \in[-1,1]$
are symmetric with respect to the imaginary axis.
The function $\varrho(x)$ in (\ref{defrho}) has
$N$ zeros in $\Im x\in [-2,2]$: 
$N/2$ zeros $\{x_j\}_{j=1}^{N/2}$
are in the physical strip $\Im x\in
[-1,1]$ and the 
others $\{x^{\prime}_j\}_{j=1}^{N/2}$ 
are out of the strip.
As $\varrho(x)$ has a property
\be
\varrho(x+2i)|_{\mu}=(-1)^{\frac{N}{2}}
\varrho(x)|_{-\mu},
\label{prorho}
\end{equation}
we have 
\be
x^{\prime}_j |_{\mu}=x_j|_{-\mu}+2i.
\label{pro}
\end{equation}  
{}From the BAE (\ref{eq.bae}),
$\{x_j^{(1)}\}$ are completely equivalent to 
$\{x_j\}$.
Thereby
one can shows from (\ref{freeT}) that $\Lambda_1(x)$
does not possess any zeros in the physical strip.

Since the function $\widetilde{\Lambda}_1(x)$ is ANZC 
$\Im x\in[-1,1]$, we have
\be
\ln \widetilde\Lambda_1(x)
= \left[K*\ln \left(X+
                    X^{-1}+2\cosh(\beta\mu)\right)\right](x). 
\end{equation}
Using the relations
\bea
\lim_{N\to\infty}\psi(x)&=&\exp\left(
             \frac{\beta}{\cosh \frac{\pi x}{2}}\right),\nn \\
\lim_{N\to\infty}\widetilde{\Lambda}(0)&=&\lim_{N\to\infty}
                                               \Lambda(0),
\eea
we obtain the free energy per site $f$ as
\bea 
f&=&-\frac{1}{\beta} \lim_{N \to \infty} 
      \ln\left(\Lambda_1(0)\right) \nonumber \\
 &=&-\frac{1}{\pi\beta} \int_0^{\frac{\pi}{2}}\ln\left(
2\cosh(\beta \cos\zeta)+
2\cosh(\beta \mu)\right) d\zeta, \label{fffe}
\eea
in agreement with Ref.~\onlinecite{Kat}.  

Next we consider the correlation length $\xi$ for
$\langle c_j^{\dagger}c_i\rangle$. 
The BAE roots $\{x^{(2)}_j\}_{j=1}^{N/2-1}$
relevant to the second largest eigenvalue
are identical with $\{x^{(1)}_j\}_{j=1}^{N/2}$
except that the largest 
magnitude one $x^{(1)}_{N/2}=\theta$ is absent.
Then the $\Lambda_2(x)$ possesses
the additional zero $\theta$ in the physical strip.
In the Trotter limit, $\theta$ is given by
\be
 \theta=\frac{2}{\pi}\sinh^{-1}
        \left(\frac{\beta}{\pi-i \beta \mu}\right).
\end{equation}  
The corresponding zero $\theta^{\prime}$ 
through the property (\ref{pro}) is
\be
\theta^{\prime}=
\frac{2}{\pi}\sinh^{-1}
        \left(\frac{\beta}{\pi+i \beta \mu}\right)+2i.
\end{equation}
The zero $\theta$ ($\theta^{\prime}$)
never goes over (never comes into)
the physical strip (see FIG.~\ref{fzeros}).
\begin{figure}
\begin{center}
\includegraphics[width=0.48\textwidth]{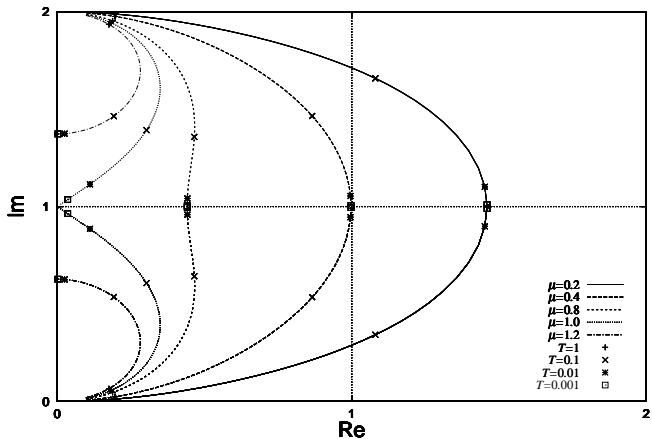}
\end{center}
\caption{The trajectories
of the additional zeros for the
free Fermion model.
The lower zero $\theta$ never goes over 
the physical strip, while
upper one $\theta^{\prime}$ never comes into
the strip. In the case $\mu=1.0$,
both zeros moves to $\theta,\theta'=i$ at the low
temperature limit.}
\label{fzeros}
\end{figure}
Consequently, $\Lambda_2(x)$ can be expressed as 
\be
\left|\Lambda_2(x)\right| =\left|\Lambda_1(x) 
\tanh\frac{\pi}{4}(x-\theta)\right|.
\end{equation}
Thus we have correlation length
$\xi$ for  
$\langle c_j^{\dagger}c_i\rangle$ as
\bea
\frac{1}{\xi}&=&-\ln\left|\tanh\left(\frac{1}{2}
        \sinh^{-1}\left(\frac{\beta}{\pi-i\beta \mu}\right)\right)
        \right| 
\nonumber \\
     &=&\frac{1}{2}\left(
       \sinh^{-1}\left(\frac{\pi+i\beta \mu}{\beta}\right)+
       \sinh^{-1}\left(\frac{\pi-i\beta \mu}{\beta}\right)
       \right).\nn \\
\label{ffxi2}
\eea
In FIG.~\ref{sfmcl2} we plot the results (\ref{ffxi2}) for some
fixed particle densities.
\begin{figure}
\begin{center}
\includegraphics[width=0.48\textwidth]{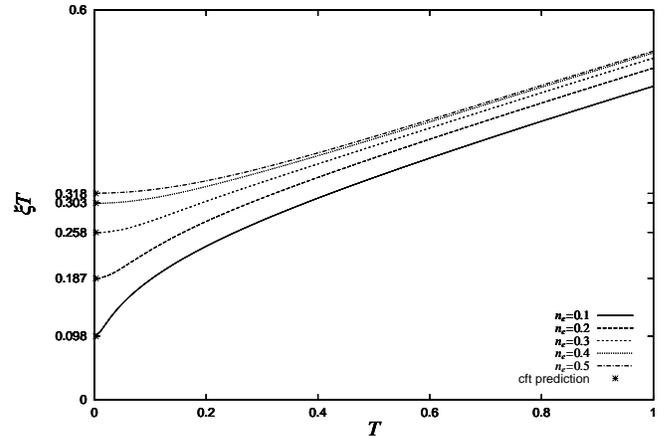}
\end{center}
\caption{The temperature dependence of
the correlation length for the
free Fermion model.}
\label{sfmcl2}
\end{figure}

\end{multicols}
\end{document}